\documentclass[aps,prb,twocolumn,english,showpacs,superscriptaddress,amssymb,amsfonts]{revtex4}
\usepackage[T1]{fontenc}
\usepackage[latin9]{inputenc}
\usepackage{amsmath}
\usepackage{dsfont}
\usepackage{amstext}

\usepackage{tocvsec2}

\usepackage{amssymb}
\usepackage{amsbsy}
\usepackage{amsthm}
\usepackage{epsfig}
\usepackage{graphicx}
\usepackage{bbm}
\usepackage{hyperref}
\usepackage{color}
\usepackage{multirow}

\newcommand{\Ref}[1]{Ref.~\onlinecite{#1}}

\newcommand{\bst}{{\boldsymbol{T}}}
\newcommand{\bsc}{{\boldsymbol{C}}}
\newcommand{\bsq}{{\boldsymbol{Q}}}
\newcommand{\bsi}{{\boldsymbol{I}}}

\newcommand{\ie}{{\emph{i.e.~}}}
\makeatletter

\newcommand{\Rmnum}[1]{\expandafter\@slowromancap\romannumeral #1@}
\makeatother
\newcommand{\imth}{\hspace{1pt}\mathrm{i}\hspace{1pt}}

\newcommand{\eg}{{\emph{e.g.~}}}

\newcommand{\mbz}{{\mathbb{Z}}}
\newcommand{\bea}{\begin{eqnarray}}
\newcommand{\eea}{\end{eqnarray}}
\newcommand{\bpm}{\begin{pmatrix}}
\newcommand{\epm}{\end{pmatrix}}
\newcommand{\bal}{\begin{aligned}}
\newcommand{\eal}{\end{aligned}}

\makeatother

\usepackage{babel}
\begin{document}
\title{Inversion symmetry protected topological insulators and superconductors}

\author{Yuan-Ming Lu}
\author{Dung-Hai Lee}
\affiliation{Department of Physics, University of California, Berkeley, California 94720, USA}
\affiliation{Materials Science Division, Lawrence Berkeley National Laboratories, Berkeley, California 94720, USA}

\begin{abstract}
Three dimensional topological insulator represents a class of novel quantum phases hosting robust gapless boundary excitations, which is protected by global symmetries such as time reversal, charge conservation and spin rotational symmetry. In this work we systematically study another class of topological phases of weakly interacting electrons protected by spatial inversion symmetry, which generally don't support stable gapless boundary excitations. We classify these inversion-symmetric topological insulators and superconductors in the framework of K-theory, and construct their lattice models. We also discuss quantized response functions of these inversion-protected topological phases, which serve as their experimental signatures.  
\end{abstract}


\maketitle



\section{Introduction}

The discovery of topological insulators\cite{Hasan2010,Hasan2011} and superconductors\cite{Qi2011} reveals a large class of novel quantum phases which, in spite of a finite energy gap for all bulk excitations, exhibit protected surface states robust against perturbations as long as certain symmetries (such as time reversal and spin rotational symmetry) are preserved. Existence of topological insulators/superconductors also discloses a rich topology of quantum phases in the presence of symmetries\cite{Chen2013}: \ie topological insulators/superconductors and the ``trivial'' ones (with no surface states) only become distinct quantum phases when certain symmetries are present. For example, a time reversal invariant (TRI) topological insulator in three dimensions\cite{Fu2007,Moore2007,Roy2009} (3d) can be continuously tuned into a 3d trivial insulator without closing the bulk energy gap, once time-reversal-breaking magnetic orders are introduced into the system\cite{Essin2009}. In general the topology of insulators/superconductors with a certain symmetry group is captured by an Abelian group, which can be (see TABLE \ref{tab:10-fold way} and \ref{tab:I square=-1} for examples) \eg a trivia group $0$ with only identity element, meaning that there is no topological insulators/superconductors; a cyclic group $\mbz_2$, meaning there is only one kind of nontrivial topological insulator/superconductor; or an integer group $\mbz$, meaning that there is an infinite number of distinct topological insulators/superconductors labeled by an integer index $\nu\in\mbz$. The addition of two elements in this Abelian group is physically realized by coupling two quantum phases together (in a way preserving symmetries) to form a new system\cite{Kitaev2009,Teo2010,Chen2013}. To understand the topology of insulators/superconductors in the presence of certain symmetries, \ie to classify different quantum phases, we need to identify the corresponding Abelian group structure of topological insulators/superconductors. Further more, physical pictures of these topological/insulators with certain symmetries can be obtained by constructing microscopic models which describes the ``root'' state of topological insulators/superconductors (mathematically the generator of this Abelian group).

The topology of non-interacting electrons with arbitrary global symmetries, such as time reversal, $U(1)$ charge conservation and $SU(2)$ spin rotational symmetry, are fully classified by the so-called ``10-fold way'' periodic table\cite{Schnyder2008,Kitaev2009} of topological insulators/superconductors. The mathematical structure behind this classification is K-theory\cite{Kitaev2009}. Intuitively the quantum critical point between two topologically distinct insulators/superconductors can in general be described by a Dirac Hamiltonian of free fermions, and the problem of classifying different insulators/superconductors in the presence of certain symmetries reduce to a mathematical problem of classifying distinct symmetry-allowed mass matrices for the Dirac fermions. Amazingly no matter what the symmetry group is, the classification (\ie the Abelian group) of distinct insulators/superconductors preserving this symmetry always falls into one of the 10-fold way. Although the topology of (non-interacting fermion) insulators/superconductors with any global symmetries is fully resolved, less is known about spatial (crystal) symmetries beyond translations\cite{Ran2010}.  Recently a lot of progress has been made along the line of topological insulators/superconductors with mirror reflection symmetry\cite{Zhang2013a,Chiu2013,Morimoto2013,Lu2014}, and 2d topological phases with crystal symmetries\cite{Fang2012,Teo2013,Benalcazar2013} (such as $C_n$ crystal rotations).

This work aims to answer the following question: how does spatial inversion symmetry reshape the topology of different insulators/superconductors with certain global symmetries? Can inversion symmetry gives rise to new topological phases\cite{Turner2010}? It turns out K-theory also provides a natural framework to classify distinct topological phases with both global symmetries and spatial inversion symmetry $\bsi$. In particular for the usual spatial inversion operation satisfying $\bsi^2=+1$, classification of topological phases protected by global symmetries and inversion $\bsi$ is summarized in TABLE \ref{tab:10-fold way}.

The presence of an extra inversion symmetry $\bsi$ modifies the original 10-fold way classification (with just global symmetries) in two fashions. First of all, many new topological phases appear, which will become trivial (\ie they will be adiabatically tuned into a trivial phase) as soon as inversion symmetry $\bsi$ is broken. Examples of this type include \eg magnetic insulators and superconductors (class A and AI), TRI insulators (class AII) and 3d singlet superconductors (class C). We explicitly construct lattice models to realize all these new topological insulators/superconductors in section \ref{sec:examples}. Secondly, certain topological phases protected by only global symmetries must break inversion symmetry $\bsi$. Therefore once we require inversion symmetry, these topological phases are forbidden. Examples of this type include TRI triplet superconductors (class DIII) and 3d TRI singlet superconductors (class CI). Their classification becomes trivial (``0'' in TABLE \ref{tab:10-fold way}) in the presence of an additional inversion symmetry $\bsi$.

We also consider a ``special'' inversion symmetry $\bsi$ satisfying $\bsi^2=-1$, \ie the square of inversion operation equals the fermion number parity in the system. Such a special inversion can be realized in 1d/2d electrons with spin-orbit coupling, by $C_2$ crystal rotation symmetry along an axis perpendicular to the 1d/2d system. It can also be realize by fermionic spinons in symmetric $Z_2$ spin liquids, which transform projectively\cite{Wen2002} under inversion symmetry $\bsi$.

This paper is organized as the following. In section \ref{sec:classification} we introduce K-theory to classify different topological phases of non-interacting electrons with both global symmetries and spatial inversion symmetry $\bsi$. A brief review of K-theory classification of gapped non-interacting fermion phases is given in Appendix \ref{app:10-fold-way}. In section \ref{sec:examples} we construct microscopic tight-binding models for the new topological insulators/superconductors protected by inversion symmetry, as summarized in TABLE \ref{tab:10-fold way}. Finally in section \ref{sec:summary} we discuss quantized responses of these inversion-symmetric topological phases as their experimental signatures and give concluding remarks.

\begin{table*}[tb]
\centering
\begin{ruledtabular}
\begin{tabular}{ |c|c|c|c|c|c|c|c|c|c|c|}
\hline
\multirow{2}{0.7cm}{\centering{AZ\\Class}}&\multirow{2}{1.3cm}{\centering{Symmetry\\group}}&{$d=1$}
&{$d=2$}&{$d=3$}&{$d=4$}&{$d=5$}&{$d=6$}&{$d=7$}&{$d=8$}
&Physical realizations\\
&&&&&&&&&&\\
\hline
A&$U(1)$&$\mbz$&$\mbz^2$&$\mbz$&$\mbz^2$&$\mbz$&$\mbz^2$&$\mbz$&$\mbz^2$&\multirow{2}{7cm}{\centering Insulators (conserving charge),\\$S^z$-conserving magnetic superconductors.}\\
&&&&&&&&&&\\
\hline
AIII&$U(1)_{spin}\times Z_2^\bst$&0&0&0&0&0&0&0&0&\multirow{2}{7cm}{\centering{TRI $S^z$-conserving superconductors.}}\\
&&&&&&&&&&\\
\hline
\hline
\hline
AI&$U(1)\rtimes Z_2^\bst$&$\mbz$&$\mbz$&$\mbz$&$\mbz^2$&$\mbz$&$\mbz$&$\mbz$&$\mbz^2$&\multirow{2}{7cm}{\centering{Insulators with a combination of time reversal and $\pi$-spin-rotation,\\$S^z$-conserving superconductors with a combination of time reversal and $\pi$-spin-rotation along $S^{x/y}$.}}\\
&$(\bst^2=+1)$&&&&&&&&&\\
&&&&&&&&&&\\
&&&&&&&&&&\\
\hline
BDI&$Z_2^\bst$&$\mbz_2$&0&0&0&0&0&$\mbz_2$&$(\mbz_2)^2$&\multirow{2}{7cm}{\centering{Superconductors with a combination of time reversal and $\pi$-spin-rotation.}}\\
&$(\bst^2=+1)$&&&&&&&&&\\
\hline
D&$Z_2^f$=N/A&$\mbz_2$&$\mbz$&0&0&0&$\mbz$&$\mbz_2$&$(\mbz_2)^2$&\multirow{2}{7cm}{\centering{Superconductors with no symmetry.}}\\
&&&&&&&&&&\\
\hline
DIII&$Z_2^\bst$&0&0&0&0&0&0&0&0&\multirow{2}{7cm}{\centering{TRI superconductors.}}\\
&$(\bst^2=-1)$&&&&&&&&&\\
\hline
AII&$U(1)\rtimes Z_2^\bst$&$\mbz$&$\mbz$&$\mbz$&$\mbz^2$&$\mbz$&$\mbz$&$\mbz$&$\mbz^2$&\multirow{2}{7cm}{\centering{TRI insulators.}}\\
&$(\bst^2=-1)$&&&&&&&&&\\
\hline
CII&$SU(2)\times Z_2^\bst$&0&0&$\mbz_2$&$(\mbz_2)^2$&$\mbz_2$&0&0&0&\multirow{2}{7cm}{\centering{Singlet superconductors with a combination of time reversal and $\pi$-pseudospin-rotation.}}\\
&$(\bst^2=+1)$&&&&&&&&&\\
\hline
C&$SU(2)_{spin}$&0&$\mbz$&$\mbz_2$&$(\mbz_2)^2$&$\mbz_2$&$\mbz$&0&0&\multirow{2}{7cm}{\centering{Singlet superconductors.}}\\
&&&&&&&&&&\\
\hline
CI&$SU(2)_{spin}\times Z_2^\bst$&0&0&0&0&0&0&0&0&\multirow{2}{7cm}{\centering{TRI singlet superconductors.}}\\
&$(\bst^2=-1)$&&&&&&&&&\\
\hline
 \end{tabular}
\caption{Classification of gapped non-interacting fermion phases with various global symmetries\cite{Schnyder2008,Kitaev2009}, in the presence of an additional inversion symmetry $\bsi$ with $\bsi^2=1$. Note that inversion symmetry $\bsi$ commutes with all other global symmetries. ``TRI'' is short for ``time reversal invariant''. The classification repeats itself when spatial dimension increases by 8.}
\label{tab:10-fold way}
\end{ruledtabular}
\end{table*}

\section{Ten-fold way classification with an extra inversion symmetry}\label{sec:classification}

In the K-theory approach\cite{Kitaev2009,Wen2012,Morimoto2013}, classification of distinct gapped phases with various symmetries is reduced to the following mathematical problem: what is the ``classifying space'' of symmetry-allowed mass matrix for a generic Dirac Hamiltonian preserving certain symmetries? Different gapped symmetric phases correspond to disconnected pieces of the classifying space, which cannot be connected to each other without closing the bulk energy gap. Mathematically the group structure formed by these different phases is given by the zeroth homotopy $\pi_0(\mathcal{S})$ of classifying space $\mathcal{S}$.

To be specific, writing fermion annihilation (and creation) operators $c_a^\dagger\equiv\eta_{2a-1}+\imth\eta_{2a}$ in terms of Majorana basis $\{\eta_a\}$, generally a quadratic Dirac Hamiltonian of non-interacting fermions has the following form
\bea\label{dirac ham}
H_{Dirac}=\imth(\sum_{i=1}^d\gamma_i\partial_i+M)
\eea
where $\{\gamma_i\}$ are real symmetric Dirac matrices, and $M$ is a real anti-symmetric mass matrix which anti-commutes with all Dirac matrices. In the Majorana basis, global symmetries such as time reversal and $U(1)$ charge conservation are all generated by real matrices (let's call them $\{g_\alpha\}$). In the K-theory approach, these symmetry generators $\{g_\alpha\}$ together with Dirac matrices $\{\gamma_i\}$ form a real (or complex) Clifford algebra $Cl_{p,q}$ (or $Cl_n$), as demonstrated in detail in Appendix \ref{app:10-fold-way}. The mass matrix $M$ serves as an extra generator, which together with Dirac matrices and symmetry generators form a bigger Clifford algebra $Cl_{p,q+1}$ (or $Cl_{n+1}$). Therefore the classifying space for symmetric mass matrix $M$ is determined by the extension problem of Clifford algebra $Cl_{p,q}\rightarrow Cl_{p,q+1}$ (or $Cl_n\rightarrow Cl_{n+1}$), and it is called $R_{q-p+2}$ (or $C_n$) for the extension of real (or complex) Clifford algebra. Due to Bott periodicity in classifying space $R_{a\mod8}$ (or $C_{n\mod2}$), the classification of topological insulators/superconductors is captured by a periodic table\cite{Kitaev2009}.

In the presence of inversion symmetry $\bsi$ in addition to the global symmetries summarized in TABLE \ref{tab:10-fold way}, how is the 10-fold-way classification\cite{Schnyder2008,Kitaev2009} of topological insulators/superconductors modified in different spatial dimensions? In the Majorana basis the inversion symmetry $\bsi$ is represented by a real symmetry matrix $I$ satisfying
\bea
\{I,\gamma_i\}=0,~~~[I,M]=[I,g_\alpha]=0,~~~I^2=+1.
\eea
since spatial derivative $\partial_i$ in (\ref{dirac ham}) changes sign under inversion. Notice that inversion matrix $I$ commutes with the generators of all global symmetries. Therefore one can define the following real matrix
\bea\label{U:definition}
U\equiv I\prod_{i=1}^d\gamma_i.
\eea
with
\bea\label{U:square}
U^2=(-1)^{d(d+1)/2}.
\eea
It's straightforward to see the following dichotomy in odd and even spatial dimensions:
\bea
\notag \{U,\gamma_i\}=\{U,g_\alpha\}=\{U,M\}=0,~~~~~&d=\text{odd};\\
\label{U:algebra} [U,\gamma_i]=[U,g_\alpha]=[U,M]=0,~~~~~&d=\text{even}.
\eea

\subsection{Two complex classes (class A and AIII)}

In the complex classes (\ie class A and AIII), the Dirac matrices $\{\gamma_i\}$ and symmetry generators $\{g_\alpha\}$ form a complex Clifford algebra $Cl_n$, due to a $U(1)$ symmetry generated by real anti-symmetric matrix $Q$ with $Q^2=-1$. Now in the presence of inversion symmetry $\bsi$, we need to consider a new matrix $U$ satisfying (\ref{U:algebra}).

\subsubsection{$d=$~odd}

In odd spatial dimensions, the matrix $U$ becomes a new generator in the complex Clifford algebra, since it anti-commutes with all other generators. The associated classifying space for mass matrix $M$ is determined by the new extension problem $Cl_{n+1}\rightarrow Cl_{n+2}$, and it changes from $C_n$ to $C_{n+1}$ due to the extra inversion symmetry $\bsi$. Hence in odd spatial dimensions, distinct gapped phases in class A form an integer group $\mbz=\pi_0(C_{0\mod2})$, and a trivial group $0=\pi_0(C_{1\mod2})$ in class AIII.

\subsubsection{$d=$~even}

In even spatial dimensions, matrix $U$ serves as a new symmetry which commutes with all generator of the original Clifford algebra. Note that $U^2=-1$ if $d=2\mod4$ and $U^2=+1$ if $d=0\mod4$. We can choose a basis where matrix $U$ (or $UQ$) is block diagonalized, so that $U=\sigma_z\otimes1$ in $d=0\mod4$ dimensions (or $UQ=\sigma_z\otimes1$ in $d=2\mod4$ dimensions). Clearly no mixing term between the $U=+1$ (or $UQ=+1$) and $U=-1$ (or $UQ=-1$) subspace is allowed by inversion symmetry $\bsi$, and each subspace has the same classification as in the standard 10-fold-way. Therefore when $d=$even, distinct gapped fermion phases have a group structure $\mbz\times\mbz$ in class A, and still a trivial group structure $0=\pi_0(C_{1\mod2})$ in class AIII, as summarized in TABLE \ref{tab:10-fold way}.

\subsection{Eight real classes}

For the eight real classes in the 10-fold-way, the classifying space for mass matrix $M$ in (\ref{dirac ham}) is related to the following extension problem of real Clifford algebra: $Cl_{p,q}\rightarrow Cl_{p,q+1}$ generated by\footnote{There is a subtlety for class C, where the real Clifford algebra is generated by $\{\gamma_iQ,C,CQ\}\rightarrow\{\gamma_iQ,C,CQ,MQ\}$. However this ultimately leads to the same conclusion as other real classes. This is discussed in detail in Appendix \ref{app:10-fold-way:real}.}
\bea
\label{extension:real}\{\gamma_i,g_\alpha\}\longrightarrow\{\gamma_i,g_\alpha,M\}.
\eea
Now with inversion symmetry $\bsi$, the new matrix $U$ satisfying (\ref{U:algebra}) will change the structure of the original Clifford algebra. From (\ref{U:square}) we can see the classification of topological insulators/superconductors with inversion symmetry changes from the original 10-fold-way, depending on spatial dimension modulo 4.

\subsubsection{$d=1\mod4$}

In this case the new generator $U$ defined in (\ref{U:definition}) satisfies
\bea\notag
\{U,\gamma_i\}=\{U,g_\alpha\}=\{U,M\}=0,~~~U^2=-1.
\eea
Therefore the original extension problem (\ref{extension:real}) of real Clifford algebra $Cl_{p,q}\rightarrow Cl_{p,q+1}$ now becomes
\bea
\label{extension:real:inversion}\{\gamma_i,g_\alpha,U\}\longrightarrow\{\gamma_i,g_\alpha,U,M\}
\eea
\ie real Clifford algebra $Cl_{p,q+1}\rightarrow Cl_{p,q+2}$. Hence the classifying space for mass $M$ changes from $R_{q-p+2}$ to $R_{q-p+3}$. This means in TABLE \ref{tab:10-fold way} the $d=1$ (and $d=5$) column shifts \emph{upward} by one row for the eight real classes, compared to the original 10-fold-way with only global symmetries (but no inversion).

\subsubsection{$d=2\mod4$}\label{sec:real->complex}

Here matrix $U$ introduced in (\ref{U:definition}) commutes with all generators of the original real Clifford algebra $Cl_{p,q}$
\bea\notag
[U,\gamma_i]=[U,g_\alpha]=[U,M]=0,~~~U^2=-1.
\eea
As discussed in Appendix \ref{app:10-fold-way:complex}, this extra inversion symmetry $\bsi$ (hence matrix $U$) will reorganizes real Clifford algebra $Cl_{p,q}\rightarrow Cl_{p,q+1}$ into complex Clifford algebra $Cl_{p+q}\rightarrow Cl_{p+q+1}$. As a result the classifying space for mass matrix $M$ in eight real classes changes from $R_{q-p+2}$ to $C_{p+q}$! Since $\pi_0(C_{0\mod2})=\mbz$ and $\pi_0(C_{1\mod2})=0$, we obtain the classification of topological insulators/superconductors with inversion symmetry in $d=2$ (and $d=6$) dimensions, as shown in TABLE \ref{tab:10-fold way}.

\subsubsection{$d=3\mod4$}

Here new generator $U$ defined in (\ref{U:definition}) satisfies
\bea\notag
\{U,\gamma_i\}=\{U,g_\alpha\}=\{U,M\}=0,~~~U^2=+1.
\eea
Hence the original extension problem (\ref{extension:real}) $Cl_{p,q}\rightarrow Cl_{p,q+1}$ changes into
\bea
\notag\{\gamma_i,g_\alpha,U\}\longrightarrow\{\gamma_i,g_\alpha,U,M\}
\eea
of real Clifford algebra $Cl_{p+1,q}\rightarrow Cl_{p+1,q+1}$. Hence the classifying space for mass $M$ changes from $R_{q-p+2}$ to $R_{q-p+1}$. Consequently in TABLE \ref{tab:10-fold way} the $d=3$ (and $d=7$) column shifts \emph{downward} by one row for the eight real classes, compared to the original 10-fold-way.

\subsubsection{$d=0\mod4$}

In this case, matrix $U$ introduced by inversion symmetry $\bsi$ again commutes with all generators of the original real Clifford algebra $Cl_{p,q}$
\bea\notag
[U,\gamma_i]=[U,g_\alpha]=[U,M]=0,~~~U^2=+1.
\eea
As discussed in Appendix \ref{app:10-fold-way:complex}, since $U^2=+1$ we can always choose a basis where $U=\sigma_z\otimes1$, and all other generators of Clifford algebra are block diagonalized. In other words the subspace with $U=+1$ and the subspace with $U=-1$ can never mix in a non-interacting Hamiltonian due to inversion symmetry $\bsi$. In each subspace the symmetry-allowed mass matrix $M$ has the same classification as the original 10-fold-way. Since the two subspaces with $U=\pm1$ are completely independent, the final classification for inversion symmetric topological insulators/superconductors has a group structure which is the square of the 10-fold-way classification. This is also summarized in TABLE \ref{tab:10-fold way} in $d=4$ (and $d=8$) dimensions.

\begin{table}[tb]
\begin{tabular}{ |c|c|c|c|c|c|c|c|c|}
\hline
AZ Class&{$d=1$}
&{$d=2$}&{$d=3$}&{$d=4$}&{$d=5$}&{$d=6$}&{$d=7$}&{$d=8$}
\\
\hline
A&$\mbz$&$\mbz^2$&$\mbz$&$\mbz^2$&$\mbz$&$\mbz^2$&$\mbz$&$\mbz^2$\\
\hline
AIII&0&0&0&0&0&0&0&0\\
\hline
\hline
\hline
AI&0&0&0&$\mbz$&$\mbz_2$&$(\mbz_2)^2$&$\mbz_2$&$\mbz$\\
\hline
BDI&0&0&0&0&0&0&0&0\\
\hline
D&$\mbz$&$\mbz^2$&$\mbz$&$\mbz$&$\mbz$&$\mbz^2$&$\mbz$&$\mbz$\\
\hline
DIII&$\mbz_2$&$(\mbz_2)^2$&$\mbz_2$&0&0&0&0&0\\
\hline
AII&$\mbz_2$&$(\mbz_2)^2$&$\mbz_2$&$\mbz$&0&0&0&$\mbz$\\
\hline
CII&0&0&0&0&0&0&0&0\\
\hline
C&$\mbz$&$\mbz^2$&$\mbz$&$\mbz$&$\mbz$&$\mbz^2$&$\mbz$&$\mbz$\\
\hline
CI&0&0&0&0&$\mbz_2$&$(\mbz_2)^2$&$\mbz_2$&0\\
\hline
 \end{tabular}
\caption{Classification of gapped non-interacting fermion phases with various global symmetries\cite{Schnyder2008,Kitaev2009}, in the presence of an additional inversion symmetry $\bsi$ with $\bsi^2=-1$. We assume inversion symmetry $\bsi$ commutes with all other global symmetries. Particularly in one and two spatial dimensions with spin-orbit coupling, symmetry $\bsi$ with $\bsi^2=-1$ can be physically realized as spatial $180^o$ rotation symmetry $C_2$. Again the classification repeats itself when spatial dimension increases by 8.}
\label{tab:I square=-1}
\end{table}

\subsection{``Special'' inversion symmetry with $\bsi^2=-1$}

In the presence of an extra ``inversion'' symmetry $\bsi$ satisfying $\bsi^2=(-1)^{\hat{N}_f}$ (\ie the inversion $\bsi$ squares to be the fermion number parity), we can also apply K-theory to classify different topological insulators/superconductors for the whole 10-fold way. The calculation is completely analogous to the case with $\bst^2=+1$, as previously discussed. The resultant classification of non-interacting fermion topological phases with global symmetries and an extra inversion symmetry satisfying $\bsi^2=-1$ is summarized in TABLE \ref{tab:I square=-1}.

For two complex classes (class A and AIII) the classifications for $\bsi^2=-1$ case are exactly the same as for $\bsi^2=+1$ case. In particular no inversion-symmetric topological insulators/superconductors exist in class AIII in any spatial dimensions, while distinct inversion-symmetric insulators in class A are labeled by one (or two) integer index (indices) in odd (or even) spatial dimensions.

For four real classes, in $d$ spatial dimensions satisfying $d=1\mod4$, the classification with $\bsi^2=-1$ shifts \emph{downward} by one row compared to the original 10-fold way with no inversion symmetry, while in $d=3\mod4$ dimensions the classification shifts \emph{upward} by one row with $\bst^2=-1$. In $d=2\mod4$ spatial dimensions, the original 10-fold way classification squares in the presence of inversion symmetry $\bsi^2=-1$. In $d=0\mod4$ spatial dimensions, the presence of inversion symmetry ($\bsi^2=-1$) effectively transform the real classes into complex classes (see section \ref{sec:real->complex}) where the classification is either $\mbz$ or $0$.

For spin-$1/2$ electrons with spin-orbit coupling in one and two spatial dimensions, such an inversion symmetry with $\bsi^2=-1$ can be realized by a combination of spatial and spin rotation by $2\pi$, \ie the $C_2$ rotation along an axis perpendicular to the 1d/2d system. For example the $\mbz^2$ classification of these $C_2$-symmetric insulators in 2d was obtained in \Ref{Fang2012}. In other cases such a ``special'' inversion symmetry $\bsi$ with $\bsi^2=-1$ can be realized in \eg $Z_2$ spin liquids\cite{Wen2002} where emergent fermion spinons transform projectively under inversion symmetry $\bsi$.

\section{Examples}\label{sec:examples}

\subsection{Class A: inversion-protected insulators}

Insulators has $U(1)$ symmetry associated with charge conservation and belong to complex class A in TABLE \ref{tab:10-fold way}. When there are no inversion symmetry, the classification of insulators (class A) are given by $0$ in one spatial dimension (1d), by $\mbz$ in 2d and by $0$ in 3d. In other words all insulators in 1d and 3d are all the same, while in 2d different insulators are described by an integer (their Hall conductance in unit of $e^2/h$). In the presence of inversion symmetry $\bsi$ with $\bsi^2=+1$, different insulators are classified by $\mbz$ in 1d, by $\mbz\times\mbz=\mbz^2$ in 2d and by $\mbz$ in 3d.

\subsubsection{$d=1$}

For class A in one spatial dimension, different inversion-symmetric insulators are labeled by an integer index $\nu\in\mbz$ where $\nu=0$ denotes the trivial insulator. Meanwhile $\nu\neq0$ corresponds to nontrivial topological insulators, which cannot be continuously connected to a trivial insulator without closing the bulk energy gap or breaking inversion symmetry.

\begin{figure}
\includegraphics[width=1\columnwidth]{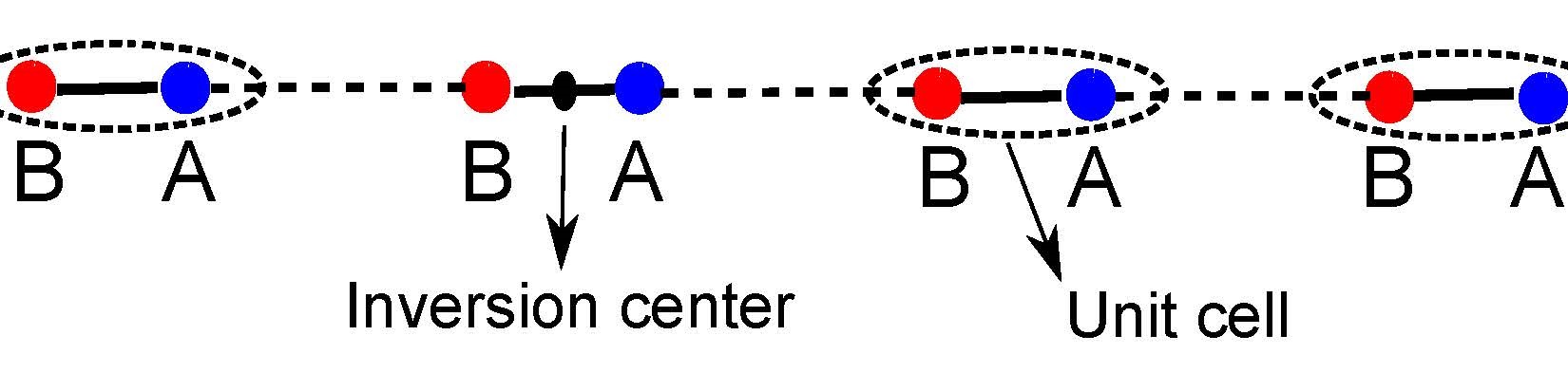}
\caption{(color online) Prototype lattice model for inversion-protected topological insulators in one spatial dimensions. Two sublattices $A$ and $B$ are labeled by blue and red solid circles and the unit cell is illustrated by the dashed oval. The inversion center lies in the middle of two sites within a unit cell. The hopping amplitude is $t_1(1+\delta)$ on solid lines and $t_1$ on dashed lines, where $t_1,\delta$ are both real parameters.}
\label{fig:1d}
\end{figure}

A simple lattice model is illustrated in FIG. \ref{fig:1d}, realized in a 1d chain of spinless (or spin-polarized) electrons with two sublattices $A$ and $B$:
\bea
&\label{ham:A:1d}\mathcal{H}^{A}_{1d}=\sum_rt_1\big[(1+\delta)c^\dagger_{r,A}c_{r,B}+c^\dagger_{r,A}c_{r+1,B}\big]+h.c.
\eea
where $t_1$ and $\delta$ are both real parameters. Without loss of generality we take $t_1>0$ as an convention in the whole paper. Clearly under inversion ${\bsi}$ we have
\bea
c_{r,A}\overset{\bsi}\longleftrightarrow c_{-r,B}
\eea
It's straightforward to see that when $\delta\ll1$ the low-energy physics is described by Dirac Hamiltonian of fermions around zone boundary $k=\pi$
\bea
&\label{dirac:A:1d}\mathcal{D}^A_{1d}=t_1\sum_k\Psi_k^\dagger\big[\delta\cdot\tau_x+k\cdot\tau_y\big]\Psi_k,\\
&\notag\Psi_k=\bpm c_{\pi+k,A}\\c_{\pi+k,B}\epm\overset{\bsi}\longrightarrow\tau_x\Psi_{-k}.
\eea
where $\vec\tau$ are Pauli matrices. When $\delta>0$ we achieve the ``root'' topological insulator protected by inversion symmetry, with topological index $\nu=1$.

\subsubsection{$d=2$}

It's well-known that irrespective of inversion symmetry, distinct insulators (class A) in two spatial dimensions are fully described by a topological invariant, \ie their Hall conductance\cite{Thouless1982} $\sigma_{xy}=C\frac{e^2}{h}$, where $C$ is an integer called Chern number. Now in the presence of an extra inversion symmetry, different insulators are characterized by a pair of integers $\vec\nu\equiv(\nu_1,\nu_2)$ where $\nu_i\in\mbz$. The Chern number is related to these two integers by
\bea
C\equiv\sigma_{xy}/(\frac{e^2}h)=\nu_1+\nu_2.
\eea

\begin{figure}
\includegraphics[width=1\columnwidth]{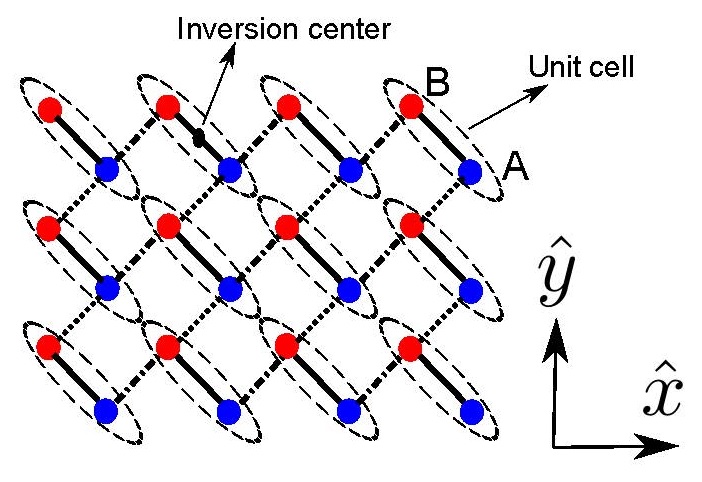}
\caption{(color online) Illustration of lattice model (\ref{ham:A:2d}) for inversion-protected topological insulators in two spatial dimensions. Two sublattices $A$ and $B$ of the checkerboard lattice are labeled by blue and red solid circles and the unit cell is illustrated by the dashed oval. The inversion center lies in the middle of two sites within a unit cell.}
\label{fig:2d}
\end{figure}

The ``root'' states of 2d inversion-symmetric insulators with $\vec\nu=(0,1)$ and $(1,0)$ can be realized by spin-$1/2$ electrons on a checkerboard lattice (with two sublattices $A$ and $B$) as illustrated in FIG. \ref{fig:2d}:
\bea
&\notag\mathcal{H}^A_{2d}=t_1\sum_{\bf r}\Big\{\big[c^\dagger_{{\bf r},A}\sigma_xc_{{\bf r}+\hat{x},B}+c^\dagger_{{\bf r},A}\sigma_zc_{{\bf r}-\hat{y},B}\\
&\notag+c^\dagger_{{\bf r},A}(\sigma_x+\sigma_z)c_{{\bf r},B}+(\delta_1-\delta_2)c^\dagger_{{\bf r},A}c_{{\bf r},B}+h.c.\big]\\
&\label{ham:A:2d}+(\delta_1+\delta_2)\sum_{s=A/B}c^\dagger_{{\bf r},s}\sigma_y c_{{\bf r},s}\Big\},\\
&\notag c_{{\bf r},s}=(c_{{\bf r},s,\uparrow},c_{{\bf r},s,\downarrow})^T,~~~s=A/B.
\eea
where Pauli matrices $\vec\sigma$ are for spin indices and $\vec\tau$ for sublattice indices. This model contains only real hoppings and ferromagnetic order along $\hat{y}$-axis. Since inversion center is located in the middle of the link between two sublattices (see FIG. \ref{fig:2d}), under inversion $\bsi$ electrons transform as
\bea
c_{{\bf r},A}\overset{\bsi}\longleftrightarrow c_{-{\bf r},B}
\eea
When $\delta_{1,2}\ll1$, the low-energy physics is governed by effective Dirac Hamiltonian around 1st BZ corner ${\bf k}=(\pi,\pi)$
\bea
&\notag\mathcal{D}^A_{2d}=t_1\sum_{\bf k}\Psi_{\bf k}^\dagger\Big\{\big[(k_x\sigma_x-k_y\sigma_z)\tau_y+\delta_1\sigma_y\big](1+\sigma_y\tau_x)\\
&\label{dirac:A:2d}+\big[(k_x\sigma_x-k_y\sigma_z)\tau_y-\delta_2\tau_x\big](1-\sigma_y\tau_x)\Big\}\Psi_{\bf k},\\
&\notag\Psi_{\bf k}\equiv\bpm c_{(\pi,\pi)+{\bf k},A}\\c_{(\pi,\pi)+{\bf k},B}\epm\overset{\bsi}\longrightarrow\tau_x\Psi_{-{\bf k}}.
\eea
The pair of integer index characterizing 2d inversion-symmetric insulators are given by
\bea
(\nu_1,\nu_2)=\big(\frac{\text{Sgn}(\delta_1)+1}2,\frac{\text{Sgn}(\delta_2)-1}2\big)
\eea
In particular it's straightforward to check that Hall conductance of the above lattice model is given by
\bea
C\equiv\sigma_{xy}/(\frac{e^2}h)=\frac{\text{Sgn}(\delta_1)}2+\frac{\text{Sgn}(\delta_2)}2.
\eea
In the pressence of inversion symmetry, two insulators with different topological index $(\nu_1,\nu_2)$ cannot be adiabatically connected to each other without closing the bulk gap.

\subsubsection{$d=3$}

\begin{figure}
\includegraphics[width=1\columnwidth]{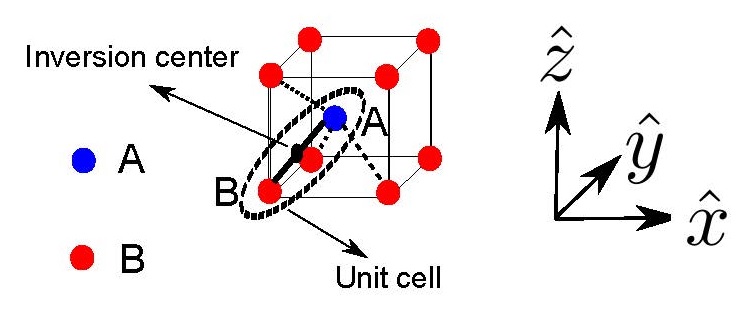}
\caption{(color online) Illustration of lattice model (\ref{ham:A:3d}) for inversion-symmetric topological insulators in three spatial dimensions. Two sublattices $A$ and $B$ of the interpenetrating primitive cubic lattice (CsCl structure) are labeled by blue and red solid circles and the unit cell is illustrated by the dashed oval. The inversion center lies in the middle of two sites within a unit cell.}
\label{fig:3d}
\end{figure}

In the absence of inversion, all insulators in three spatial dimensions are topologically the same.  As shown in TABLE \ref{tab:10-fold way}, 3d insulators protected by inversion symmetry $\bsi$ are characterized by an integer index $\nu\in\mbz$. The root state with $\nu=1$ can be realized by spin-$1/2$ electrons on an interpenetrating primitive cubic lattice (\ie CsCl structure), as shown in FIG. \ref{fig:3d}.

There are two sublattices labeled again by index $s=A/B$ and the lattice model writes
\bea
&\label{ham:A:3d}\mathcal{H}^A_{3d}=t_1\sum_{\bf r}\Big\{c^\dagger_{{\bf r},A}(\sigma_x+\sigma_y+\sigma_z+\delta)c_{{\bf r},B}\\
\notag&+c^\dagger_{{\bf r},A}\sigma_xc_{{\bf r}+\hat{x},B}+c^\dagger_{{\bf r},A}\sigma_yc_{{\bf r}+\hat{y},B}
+c^\dagger_{{\bf r},A}\sigma_zc_{{\bf r}+\hat{z},B}\Big\}+~h.c.,\\
&\notag c_{{\bf r},s}=(c_{{\bf r},s,\uparrow},c_{{\bf r},s,\downarrow})^T,~~~s=A/B.
\eea
where the two sublattices are related by inversion symmetry
\bea
c_{{\bf r},A}\overset{\bsi}\longleftrightarrow c_{-{\bf r},B}
\eea
When $\delta\ll1$ the system is described by a low-energy Dirac Hamiltonian around zone corner ${\bf k}=(\pi,\pi,\pi)$
\bea
&\label{dirac:A:3d}\mathcal{D}^A_{3d}=\sum_{\bf k}t_1\Psi_{\bf k}^\dagger({\bf k}\cdot\vec\sigma\tau_y+\delta\tau_x)\Psi_{\bf k},\\
&\notag\Psi_{\bf k}\equiv\bpm c_{(\pi,\pi)+{\bf k},A}\\c_{(\pi,\pi)+{\bf k},B}\epm\overset{\bsi}\longrightarrow\tau_x\Psi_{-{\bf k}}.
\eea
where $\vec\sigma$ and $\vec\tau$ are Pauli matrices for spin and sublattice indices. Specifically $\delta>0$ corresponds to the $\nu=1$ topological insulator, while $\delta<0$ leads to the $\nu=0$ trivial insulator which can be adiabatically connected to an atomic insulator.

\subsection{Class AI: inversion-protected magnetic insulators/superconductors}

The symmetry group for real class AI is $U(1)\rtimes Z_2^\bst$ where ani-unitary time reversal operation $\bst$ satisfies $\bst^2=+1$. It can be realized both in insulators of spin-polarized electrons and in superconductors with collinear magnetic order. In the absence of other symmetries, the classification for class AI is always trivial in 1d, 2d and 3d. When there is an extra inversion symmetry, distinct insulators emerge which are classified by an integer index $\nu\in\mbz$ in 1d, 2d and 3d.

\subsubsection{$d=1$}

Clearly the 1d lattice Hamiltonian (\ref{ham:A:1d}) of spin-polarized electrons also has time reversal symmetry $\bst$ (here $\bst^2=+1$ for spinless electrons), since all the hopping parameters are real. Therefore (\ref{ham:A:1d}) also realizes the ``root'' state for 1d inversion-symmetric topological insulators in class AI when $\delta>0$.

\subsubsection{$d=2$}

For 2d insulators in class AI, their Hall conductance (or Chern number $C$) must vanish due to time reversal symmetry $\bst$. Note that any 2d insulator with inversion symmetry is fully described by a pair of integers $(\nu_1,\nu_2)$ where Chern number is given by $C=\nu_1+\nu_2$. Therefore for time reversal invariant (TRI) insulators (class AI) with inversion symmetry, they have $\nu_1+\nu_2=0$ and are fully characterized by one integer
\bea
\nu\equiv\nu_1=-\nu_2.
\eea
Naturally the ``root'' state for 2d inversion-symmetric insulators in class AI is given by the same lattice model (\ref{ham:A:2d}) as in class A, where $\uparrow/\downarrow$ now become indices for two different \emph{orbitals} instead of for spin, and $\delta_1+\delta_2=0$ due to time reversal symmetry. Since there are real (intra-orbital and inter-orbital) hoppings in (\ref{ham:A:2d}) with $\delta_1=-\delta_2$, such a lattice model of spin-polarized two-orbital electrons preserve anti-unitary time reversal symmetry $\bst$ with $\bst^2=1$
\bea
\bpm c_{{\bf r},A}\\c_{{\bf r},B}\epm\overset{\bst}\longrightarrow\bpm c_{{\bf r},A}\\c_{{\bf r},B}\epm,~~\Psi_{\bf k}\overset{\bst}\longrightarrow\Psi_{-{\bf k}}.\notag
\eea
in (\ref{ham:A:2d}) and (\ref{dirac:A:2d}). In particular the integer topological index $\nu$ is given by
\bea
\nu=\frac{\text{Sgn}(\delta_1)+1}2=\frac{1-\text{Sgn}(\delta_2)}2.
\eea
as time reversal requires $\delta_1=-\delta_2$. So the root state with $\nu=1$ has $\delta_1=-\delta_2>0$ in (\ref{ham:A:2d}).

\begin{figure}
\includegraphics[width=1\columnwidth]{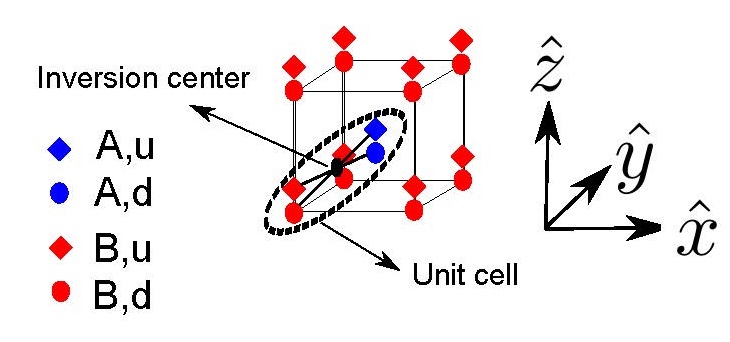}
\caption{(color online) Illustration of lattice model (\ref{ham:AI:3d}) for inversion-symmetric topological superconductors (class AI) in three spatial dimensions. Four sublattices are labeled by two indices $s=A/B$ and $f=u/d$. Index $s=A/B$ are denoted by blue/red colors, index $f=u/d$ by diamond/circle, and the unit cell is illustrated by the dashed oval. The inversion center lies in the crossing of the link between $(A,u)\leftrightarrow(B,d)$ and the one between $(A,d)\leftrightarrow(B,u)$.}
\label{fig:3d_AI}
\end{figure}

\subsubsection{$d=3$}

Inversion-symmetric topological phases in class AI can be realized by $S^z$-conserving superconductors of spin-$1/2$ electrons in a 3d lattice illustrated by FIG. \ref{fig:3d_AI}, where there are 4 sublattices per unit cell labeled by two indices $s=A/B$ and $f=u/d$. The anti-unitary $\bst$ symmetry is the combination of usual time reversal operation and $S^y$ spin rotation by angle $\pi$, so that electron spin doesn't flip under $\bst$ operation. The lattice model is given by
\bea
&\notag\mathcal{H}_{3d}^{AI}=t_1\sum_{\bf r}\Big\{\psi^\dagger_{\bf r}\big[\sigma_x+\sigma_z+(1+\delta)\mu_x\big]\tau_x \psi_{\bf r}\\
&\notag+(\psi^\dagger_{{\bf r},A}\sigma_x \psi_{{\bf r}+\hat{x},B}+\psi^\dagger_{{\bf r},u}\tau_x \psi_{{\bf r}+\hat{y},d}\\
&\label{ham:AI:3d}+\psi^\dagger_{{\bf r},A}\sigma_z\psi_{{\bf r}+\hat{z},B}+h.c.)\Big\},\\
&\notag c_{\bf r}\equiv\bpm c_{{\bf r},A,u}\\c_{{\bf r},A,d}\\c_{{\bf r},B,u}\\c_{{\bf r},B,d}\epm\overset{\bsi}\longrightarrow\bpm c_{-{\bf r},B,d}\\c_{-{\bf r},B,u}\\c_{-{\bf r},A,d}\\c_{-{\bf r},A,u}\epm=\tau_x\mu_x c_{-{\bf r}},\\
&\notag c_{\bf r}\overset{\bst}\longrightarrow c_{\bf r},~~~\psi_{\bf r}\equiv\bpm c_{{\bf r},\uparrow}\\c^\dagger_{{\bf r},\downarrow}\epm
\eea
where Pauli matrices $\vec\sigma$ are for Nambu index $n=c_\uparrow/c^\dagger_\downarrow$, $\vec\tau$ for $s=A/B$ and $\vec\mu$ for $f=u/d$. When $\delta\ll1$ the low-energy physics is governed by Dirac fermions around zone corner ${\bf k}=(\pi,\pi,\pi)$:
\bea
&\notag\mathcal{D}_{3d}^{AI}=t_1\sum_{\bf k}\Psi^\dagger_{\bf k}\big[(k_x\sigma_x+k_z\sigma_z)\tau_y+k_y\tau_x\mu_y+\delta\cdot\tau_x\mu_x\big]\Psi_{\bf k},\\
&\label{dirac:AI:3d}\Psi_{\bf k}\equiv \bpm c_{(\pi,\pi,\pi)+{\bf k},\uparrow}\\c^\dagger_{(\pi,\pi,\pi)-{\bf k},\downarrow}\epm\overset{\bsi}\longrightarrow\tau_x\mu_x\Psi_{-{\bf k}},\\
&\notag\Psi_{\bf k}\overset{\bst}\longrightarrow\Psi_{-{\bf k}}.
\eea
Here $\delta>0$ gives rise to the nontrivial $\nu=1$ inversion-protected insulator in class AI, while $\delta<0$ leads to the trivial $\nu=0$ insulator which can be continuously tuned into an atomic insulator without closing energy gap.

\subsection{Class AII: inversion-protected TRI insulators}

Class AII has symmetry group $U(1)\rtimes Z_2^\bst$ where $\bst^2=-1$. It describes insulators of spin-$1/2$ electrons with time reversal symmetry obeying $\bst^2=-1$. As usual spin-$1/2$ electrons transform as
\bea
c_{\bf r}\equiv\bpm c_{{\bf r},\uparrow}\\c_{{\bf r},\downarrow}\epm\overset{\bst}\longrightarrow\bpm c_{{\bf r},\downarrow}\\-c_{{\bf r},\uparrow}\epm=\imth\sigma_y c_{\bf r}
\eea
Without any extra symmetry, different TRI spin-$1/2$ insulators (class AII) are classified by $0$ in 1d and $\mbz_2$ in 2d, 3d. This means all TRI spin-$1/2$ insulators are trivial in 1d, but in 2d and 3d there is one type of nontrivial insulator\cite{Hasan2010,Hasan2011,Qi2011}: \ie 2d quantum spin Hall insulator (QSHI) in 2d and 3d topological insulator. When we further consider inversion symmetry $\bsi$ with $\bsi^2=+1$, the classification for different class AII insulators becomes $\mbz$ in $d=1,2,3$. In other words, in one, two and three spatial dimensions, distinct inversion-symmetric insulators in class AII are labeled by an integer $\nu\in\mbz$.

\subsubsection{$d=1$}

The root state of 1d inversion-symmetric topological insulator in class AII can be easily realized by spin-$1/2$ electrons on a 1d chain with inversion center illustrated in FIG. \ref{fig:1d}, where spin-$\uparrow$ electrons and spin-$\downarrow$ electrons have the same real hopping terms as described in (\ref{ham:A:1d}). Again $\delta>0$ leads to the nontrivial $\nu=1$ insulator, which is robust against any perturbation as long as time reversal, inversion and $U(1)$ charge conservation are preserved. $\delta<0$ corresponds to the trivial insulator.

\subsubsection{$d=2$}

Here due to time reversal, the Chern number or Hall conductance has to vanish for any 2d insulator in class AII. Regardless of inversion symmetry, there is already one nontrivial 2d topological insulator in class AII, \ie the $\mbz_2$ quantum spin Hall insulator\cite{Kane2005,Kane2005a,Bernevig2006}. QSHI preserving inversion symmetry can be realized by spin-$1/2$ electrons on a checkerboard lattice (in FIG. \ref{fig:2d}):
\bea
&\notag\mathcal{H}_{2d}^{AII}=t_1\sum_{\bf r}\Big[\big(c^\dagger_{\bf r}\frac{\tau_x-\imth\sigma_x\tau_z}2c_{{\bf r}+\hat{x}}+
c^\dagger_{\bf r}\frac{\tau_x-\imth\sigma_y\tau_z}2c_{{\bf r}+\hat{y}}\\
&+h.c.\big)\label{ham:AII:2d}+(2+\delta)c_{\bf r}^\dagger\tau_x c_{\bf r}\Big],\\
&c_{{\bf r},A}\overset{\bsi}\longleftrightarrow c_{-{\bf r},B}~~\text{or}~~c_{\bf r}\overset{\bsi}\longrightarrow\tau_x c_{-{\bf r}}.\notag
\eea
where $\vec\sigma$ are Pauli matrices for spin indices $\uparrow/\downarrow$ and $\vec\tau$ for sublattice index $s=A/B$. When real parameter $\delta\ll1$ the low-energy physics is described by Dirac fermions around zone corner ${\bf k}=(\pi,\pi)$:
\bea
&\label{dirac:AII:2d}\mathcal{D}_{2d}^{AII}=\sum_{\bf k}\Psi^\dagger_{\bf k}\big[(k_x\sigma_x+k_y\sigma_y)\tau_z+\delta\tau_x\big]\Psi_{\bf k},\\
&\notag\Psi_{\bf k}=c_{(\pi,\pi)+{\bf k}}\overset{\bst}\longrightarrow\imth\sigma_y\Psi_{-{\bf k}},\\
&\notag\Psi_{\bf k}\overset{\bsi}\longrightarrow\tau_x\Psi_{-{\bf k}}.
\eea
Real parameter $\delta>0$ leads to a nontrivial ``root'' topological insulator with $\nu=1$, while $\delta<0$ leads to the trivial $\nu=0$ state. In the absence of inversion symmetry, once two layers of $\nu=1$ state are stacked together, the resulting $\nu=2$ state can be adiabatically deformed into a trivial atomic insulator ($\nu=0$) without closing the bulk energy gap. When there is inversion symmetry, however, $\nu=2$ state cannot be deformed into trivial $\nu=0$ state without going through a phase transition. Instead there are an integer number (labeled by $\nu$) of different 2d class AII insulators protected by inversion symmetry.

\subsubsection{$d=3$}

Irrespective of inversion symmetry, in three spatial dimensions there is already a nontrivial topological insulator in class AII: the 3d topological insulator\cite{Fu2007,Moore2007,Roy2009}. A simple lattice model for 3d TRI topological insulators of spin-$1/2$ electrons can be realized on a CsCl-like lattice (with two sublattices with index $s=A/B$) as shown in FIG. \ref{fig:3d}
\bea
&\notag\mathcal{H}_{3d}^{AII}=t_1\sum_{\bf r}\Big\{\big(c^\dagger_{\bf r}\frac{\tau_x-\imth\sigma_x\tau_z}2c_{{\bf r}+\hat{x}}+c^\dagger_{\bf r}\frac{\tau_x-\imth\sigma_y\tau_z}2c_{{\bf r}+\hat{y}}\\
&+c^\dagger_{\bf r}\frac{\tau_x-\imth\sigma_z\tau_z}2c_{{\bf r}+\hat{z}}+h.c.\big)\label{ham:AII:3d}+(\delta+3)c^\dagger_{\bf r}\tau_x c_{\bf r}\Big\},\\
&\notag c_{{\bf r}}\overset{\bst}\longrightarrow\imth\sigma_y c_{\bf r},\\
&\notag c_{{\bf r},A}\overset{\bsi}\longleftrightarrow c_{-{\bf r},B}~~\text{or}~~c_{\bf r}\overset{\bsi}\longrightarrow\tau_x c_{-{\bf r}}.
\eea
where Pauli matrices $\vec\sigma$ are for spin index $\uparrow/\downarrow$, and $\vec\tau$ for sublattice index $s=A/B$. Note that the above model also preserves inversion symmetry $\bsi$.

When real parameter $\delta\ll1$, Dirac fermions around zone corner ${\bf k}=(\pi,\pi,\pi)$ describe the low-energy physics of the system
\bea
&\mathcal{D}_{3d}^{AII}=\sum_{\bf k}\Psi^\dagger_{\bf k}({\bf k}\cdot\vec\sigma\tau_z+\delta\tau_x)\Psi_{\bf k},\\
&\notag\Psi_{\bf k}\equiv c_{(\pi,\pi,\pi)+{\bf k}}\overset{\bst}\longrightarrow\imth\sigma_y\Psi_{-{\bf k}},\\
&\notag\Psi_{\bf k}\overset{\bsi}\longrightarrow\tau_x\Psi_{-{\bf k}}.
\eea
Here in this model $\delta>0$ leads to the nontrivial $\nu=1$ topological insulator (class AII) in 3d, while $\delta<0$ leads to the trivial $\nu=0$ insulator which doesn't have protected Dirac cones on the surface.

Without inversion symmetry, different class-AII insulators have a $\mbz_2$ classification in the sense that two copies of $\nu=1$ topological insulators, when coupled together, can be continuously tuned into a trivial $\nu=0$ insulator without closing bulk energy gap. In the presence of inversion symmetry $\bsi$, on the other hand, such a adiabatic deformation is not possible and there are an integer number of distinct insulators in class AII, labeled by integer index $\nu\in\mbz$.

\subsection{Class C: inversion-protected singlet superconductors in $d=3$}

In the absence of inversion symmetry, there are no topological singlet superconductors (class C) in 3d, \ie any two singlet superconductors in three dimensions can be continuously deformed into each other with phase transitions. However, an extra inversion symmetry gives rise to an integer classification, \ie distinct inversion-symmetric singlet superconductors are labeled by an integer $\nu\in\mbz$.

The $\nu=1$ ``root'' state can be realized on by two-orbital spin-$1/2$ electrons on a interpenetrating primitive cubic lattice (CsCl structure, see FIG. \ref{fig:3d}):
\bea
&\notag\mathcal{H}_{3d}^{C}=t_1\sum_{\bf r}\Big\{\psi^\dagger_{{\bf r}}\big[2(\sigma_x+\sigma_y+\sigma_z)+\delta\cdot\mu_y\big]\tau_x \psi_{\bf r}\\
&\notag+(\psi^\dagger_{{\bf r},A}\sigma_x \psi_{{\bf r}+\hat{x},B}+\psi^\dagger_{{\bf r},u}\sigma_y \psi_{{\bf r}+\hat{y},d}\\
&\label{ham:C:3d}+\psi^\dagger_{{\bf r},A}\sigma_z\psi_{{\bf r}+\hat{z},B}+h.c.)\Big\},\\
&\notag c_{\bf r}\equiv\bpm c_{{\bf r},A}\\c_{{\bf r},B}\epm\overset{\bsi}\longrightarrow\bpm c_{-{\bf r},B}\\c_{-{\bf r},A}\epm=\tau_xc_{-{\bf r}},\\
&\notag c_{\bf r}\overset{e^{\imth\pi S^y}}\longrightarrow\imth\sigma_yc_{\bf r},~~~\psi_{\bf r}\equiv\bpm c_{{\bf r},\uparrow}\\c^\dagger_{{\bf r},\downarrow}\epm.
\eea
where Pauli matrices $\vec\sigma$ are for Nambu index $c_\uparrow/c^\dagger_{\downarrow}$, $\vec\tau$ for sublattice index $s=A/B$ and $\vec\mu$ for orbital index.

When $\delta\ll1$ we obtain an effective Dirac Hamiltonian around zone corner ${\bf k}=(\pi,\pi,\pi)$
\bea
&\label{dirac:C:3d}\mathcal{D}_{3d}^C=t_1\sum_{\bf k}\Psi^\dagger_{\bf k}(2{\bf k}\cdot\vec\sigma\tau_y+\delta\cdot\tau_x\mu_y)\Psi_{\bf k},\\
&\notag\Psi_{\bf k}\equiv\bpm c_{(\pi,\pi,\pi)+{\bf k},\uparrow}\\c^\dagger_{(\pi,\pi,\pi)-{\bf k},\downarrow}\epm\overset{e^{\imth\pi S^y}}\longrightarrow\imth\sigma_y\Psi_{-{\bf k}}^\dagger,\\
&\notag\Psi_{\bf k}\overset{\bsi}\longrightarrow\tau_x\Psi_{-{\bf k}}.
\eea
Here $\delta>0$ leads to the nontrivial $\nu=1$ topological singlet superconductor in 3d, while $\delta<0$ corresponds to the trivial $\nu=0$ superconductor which can be continuously tuned into a ``strong-pairing'' $s$-wave superconductor\cite{Read2000}.

\section{Discussions}\label{sec:summary}

\subsection{Quantized response of inversion-protected topological insulators and superconductors}

Unlike topological phases protected by global symmetries, these inversion-protected topological phases cannot be identified by their gapless boundary states, simply because the boundary itself necessarily breaks inversion symmetry\cite{Hughes2011,Turner2010}. A natural question is: are there any experimental observable for these inversion-protected topological insualtors/superconductors? The answer is yes. For example it was pointed out in \Ref{Turner2012} that inversion-protected topological insulators (class A) can be
diagnosed by its electromagnetic response: specifically the $\nu=1$ inversion-protected class-A insulator in 3d, realized in lattice model (\ref{ham:A:3d}), exhibits the so-called topological magneto-electric effect\cite{Qi2008,Essin2009}:
\bea
{\bf M}=\theta\frac{e^2}{2\pi h}{\bf E},~~~\theta=\pi\mod2\pi.
\eea
\ie an applied electric field ${\bf E}$ will induced a magnetization ${\bf M}$ proportional to the electric field, where the proportionality constant is quantized. Another equivalent manifestation is the ``half'' quantum Hall effect on the gapped surface of inversion-symmetric $\nu=1$ topological insulators in class A. More precisely, the (charge) Hall conductance of gapped surface is quantized as
\bea
\sigma_{xy}=\frac{e^2}{h}\cdot\frac{\theta}{2\pi},~~~\theta=\pi\mod2\pi.
\eea
Here $\theta$ is defined modular $2\pi$, and it changes sign under either time reversal or spatial inversion operation. Therefore $\theta$ takes the value of either $0$ or $\pi$ for the inversion-protected $\nu=1$ 3d insulator in class A, AI and AII. In all these cases distinct 3d insulators are labeled by an integer index $\nu\in\mbz$, and $\theta=0$ for $\nu=$~even while $\theta=\pi$ for $\nu=$~odd. Notice that in a pure 2d insulator (preserving $U(1)$ charge conservation), the charge Hall conductance must be an integer\cite{Thouless1982} in unit of $e^2/h$.

For inversion-protected singlet superconductors in class C (labeled by an integer $\nu\in\mbz$), a similar anomaly appears on the gapped surface of $\nu=1$ topological superconductor. Although there is no $U(1)$ charge conservation, $SU(2)$ spin rotational symmetry allows a well-defined (and quantized) spin Hall conductance\cite{Senthil1999} $\sigma_{xy}^{\text{spin}}$:
\bea
j^{{\bf S}}_x=\sigma_{xy}^{\text{spin}}\cdot\big[-\frac{\text{d}{\bf B}(y)}{\text{d}y}\big]
\eea
where $j^{\bf S}$ denotes the spin current and ${\bf B}$ is the applied magnetic field. As derived in Appendix \ref{app:surface:class C:3d}, the gapped surface of $\nu=1$ singlet superconductor in class C exhibits a ``half'' spin quantum Hall effect:
\bea
\sigma_{xy}^{\text{spin}}=\frac{\theta_s}{2\pi}\cdot\frac{\hbar}{8\pi},~~~\theta_s=\pi\mod2\pi.
\eea
This is associated to the $SU(2)$ theta term\cite{Schnyder2009} $\mathcal{L}=\frac{\theta_s}{32\pi^2}\epsilon^{\mu\nu\rho\lambda}{Tr}(\hat{F}_\mu\nu\hat{F}_{\rho\lambda})$, describing responses of inversion-protected $\nu=1$ singlet superconductor to an external $SU(2)$ ``spin gauge field'' with field strength $\hat{F}_{\mu\nu}$. Again $\theta$ is defined modular $2\pi$, and is odd under inversion operation $\bsi$. Therefore $\theta$ is quantized to be $\pi$ for all inversion-protected $\nu=$~odd states in class C, and $0$ for $\nu=$~even states in class C. In a gapped 2d system with $SU(2)$ spin rotational symmetry, the spin quantum Hall conductance must be an integer in unit of $\hbar/4\pi$, therefore such a ``half-integer'' $\sigma_{xy}^{\text{spin}}$ here is a surface anomaly of 3d inversion-protected singlet superconductor.

\subsection{Concluding remarks}

In this work we classify and construct the inversion-symmetric topological insulators/superconductors of non-interacting electrons, which cannot be adiabatically tuned (no bulk gap closing) into a trivial atomic insulator/superconductor without breaking inversion symmetry $\bsi$ or other global symmetries. Although these inversion-protected topological phases generally doesn't support robust gapless surface states, they can be partially characterized by their quantized response functions. For example, the parity of integer index $\nu\in\mbz$ labeling inversion-protected topological phases in 3d can be determined by the ``half'' charge/spin quantum Hall effect on their gapped surfaces. 

Although the topology of inversion-symmetric insulators/superconductors of non-interacting electrons cannot be fully captured by their quantized responses or surface states, entanglement spectrum in principle can provide a comprehensive diagnosis\cite{Turner2010,Hughes2011,Fang2013a}. This is because unlike a real boundary, an entanglement cut may preserve inversion symmetry. It'll be an interesting future direction to understand the inversion-protected topological phases in terms of their entanglement spectrum. Meanwhile, we've so far focused on non-interacting electron systems, where strong interactions could in principle change the topology (and classification) of topological insulators/superconductors\cite{Fidkowski2010,Turner2011} drastically. We leave these two aspects to future study. \\

\emph{Note added}--- Upon completion of this work we became aware of \Ref{Shiozaki2014}, where the same classification for non-interacting topological phases was obtained independently.

\acknowledgements

We are indebted to Ari Turner for helpful discussions. YML thanks Aspen Center for Physics for hospitality where this work was initiated. This work is supported by Office of BES, Materials Sciences Division of the U.S. DOE under contract No. DE-AC02-05CH11231 (YML,DHL) and in part by the National Science Foundation under Grant No. PHYS-1066293(YML).

\appendix

\section{K-theory classification of non-interacting topological insulators and superconductors}\label{app:10-fold-way}

In this section we briefly review the ten-fold way classification\cite{Schnyder2008,Kitaev2009} of non-interacting topological insulators and superconductors in the K-theory approach\cite{Kitaev2009,Wen2012,Morimoto2013}. The idea is to identify the ``classifying space'' of symmetry-allowed mass terms to Dirac Hamiltonians in various spatial dimensions, and disconnected pieces of the classifying space (its zeroth homotopy group $\pi_0$) correspond to topologically distinct gapped symmetric states.

\subsection{General discussions}

Let's start by representing quadratic Dirac Hamiltonians in the Majorana basis\cite{Kitaev2009} $\{\eta_a\}$, where the fermion annihilation operator is written as $c^\dagger_l=\eta_{2l-1}+\imth\eta_{2l}$. A generic Dirac Hamiltonian in $d$ spatial dimension can be written as
\bea\label{dirac}
\hat{H}_{Dirac}=\imth\sum_{a,b}\eta_a\big[\sum_{i=1}^d(\partial_i\gamma_i)_{a,b}+M_{a,b}\big]\eta_b
\eea
with real symmetric Dirac matrices $\{\gamma_i\}$ satisfying
\bea\label{cond:gamma}
\{\gamma_i,\gamma_j\}=2\delta_{i,j},~~\gamma_i^T=\gamma_i,~~\{\gamma_i,M\}=0.
\eea
The mass matrix $M$ is real and anti-symmetric. Without altering topological properties we can always flatten the spectrum by choosing
\bea\label{cond:mass}
M^2=-1,~~~M=-M^T.
\eea
In the presence of a global $U(1)$ symmetry $\bsq$, the conserved $U(1)$ charge is represented as $\hat{N}=\imth\eta_a Q_{a,b}\eta_b$ where $Q=-Q^T$ is a real anti-symmetric matrix satisfying
\bea
[Q,\gamma_i]=[Q,M]=0,~~~Q^2=-1.
\eea
In general a unitary $Z_2$ symmetry (such as particle-hole symmetry or $\pi$ spin rotation) $\bsc$ is represented by a real matrix $C$ satisfying
\bea
[C,\gamma_i]=[C,M]=0,~~~C^TC=1.
\eea
For example if $\bsq$ corresponds to charge conservation and $\bsc$ is particle-hole symmetry, we have
\bea\notag
C^2=+1,~~~C=C^T,~~~\{Q,C\}=0.
\eea
One the other hand if $\bsq$ represents $S^z$ spin conservation and $\bsc$ is $\pi$ spin rotation along $S^y$-axis, we have
\bea\notag
C^2=-1,~~~C=-C^T,~~~\{Q,C\}=0.
\eea
Meanwhile anti-unitary time reversal symmetry $\bst$ is represented by a real matrix $T$ satisfying
\bea
\{T,\gamma_i\}=\{T,M\}=0,~~~T^TT=1.
\eea
Just like unitary $Z_2$ symmetry $\bsc$, here the real matrix $T$ can be symmetric or anti-symmetric depending on $\bst^2=\pm1$:
\bea
T^2=\pm1,~~~T^T=\pm T.
\eea
In a spin-$1/2$ electronic system, the conventional time reversal symmetry has $T^2=-1$. One can also define an anti-unitary symmetry $\bst$ as a combination of time reversal and a $\pi$ spin rotation, and $T^2=+1$ for this redefined ``time reversal'' operation. $T$ may commute or anticommute with $U(1)$ symmetry generator $Q$ and $Z_2$ symmetry generator $C$, depending on the specific symmetry group.

In the K-theory approach to topological insulators/superconductors, the Dirac matrices $\{\gamma_i\}$ and symmetry operators always form a real (or complex) Clifford algebra $Cl_{p,q}$ (or $Cl_n$). The classifying space of symmetry-allowed mass matrix $M$ is determined by the extension of this Clifford algebra to $Cl_{p,q+1}$ (or $Cl_{n+1}$) by adding the mass matrix $M$ as an extra generator. The classifying space associated with the above extension problem is $R_{q-p+2\mod8}$ (or $C_{n\mod2}$) due to Bott periodicity. Distinct gapped symmetric phases are classified by disconnected pieces of the classifying space, \ie by zeroth homotopy group $\pi_0(R_{q-p+2\mod8})$ or $\pi_0(C_{n\mod2})$. The periodicity 8 (or 2) for classifying spaces is the so-called Bott periodicity. The exact form of classifying spaces $R_a$ (or $Cl_n$), as well as their zeroth homotopy group are given \eg in \Ref{Kitaev2009,Wen2012,Morimoto2013}. Notice that any 0th homotopy group $\pi_0(X)$ is always Abelian.

There are $p+q$ generators (real orthogonal matrices $\{e_i\}$) for a real Clifford algebra $Cl_{p,q}$ satisfying
\bea\label{clifford:real}
&\{e_i,e_j\}=0,~~~e_i^2=\begin{cases}+1,~~~&1\leq i\leq p,\\-1,~~~&p+1\leq i\leq p+q.\end{cases}
\eea
Meanwhile a complex Clifford algebra $Cl_n$ has $n$ generators (complex Hermitian matrices $\{e_a\}$) satisfying
\bea\label{clifford:complex}
\{e_a,e_b\}=2\delta_{a,b}.
\eea\\

\subsection{Eight real classes}\label{app:10-fold-way:real}

Let's first consider class D with no symmetry at all\footnote{We don't consider fermion parity $(-1)^{\hat{N}_f}$ as a symmetry since it cannot be broken in any local Hamiltonian.}. In $d$ spatial dimensions we consider extension problem of Clifford algebra $Cl_{d,0}\rightarrow Cl_{d,1}$ generated by
\bea
&\notag\{\gamma_1,\cdots,\gamma_d\}\longrightarrow\{\gamma_1,\cdots,\gamma_d,M\}
\eea
The associated classification of topological superconductors in class D is given by $\pi_0(R_{2-d})$ in $d$ dimensions.

Now let's consider time reversal symmetry $\bst$, \ie extension problem
\bea
&\notag\{\gamma_1,\cdots,\gamma_d,T\}\longrightarrow\{\gamma_1,\cdots,\gamma_d,T,M\}
\eea
For $\bst^2=+1$ we have extension $Cl_{d+1,0}\rightarrow Cl_{d+1,1}$, and hence topological superconductors in class BDI are classified by $\pi_0(R_{1-d})$ in $d$ dimensions. On the other hand if $\bst^2=-1$, the extension problem is $Cl_{d,1}\rightarrow Cl_{d,2}$, and hence topological superconductors in class DIII are classified by $\pi_0(R_{3-d})$ in $d$ dimensions.

Adding a $U(1)$ charge conservation symmetry (generated by $\bsq$) to time reversal $\bst$, the Clifford algebra extension problem becomes
\bea
&\notag\{\gamma_1,\cdots,\gamma_d,T,TQ\}\longrightarrow\{\gamma_1,\cdots,\gamma_d,T,TQ,M\}
\eea
Notice that $\{T,Q\}=0$ here. Again depending on the sign of $\bst^2=\pm1$, the classification of topological insulators with time reversal symmetry is given by $\pi_0(R_{-d})$ for class AI ($\bst^2=+1$), and by $\pi_0(R_{4-d})$ for class AII ($\bst^2=-1$).

If we consider a spin rotational $U(1)$ symmetry $\bsq$ (say, along $S^z$-axis) instead, it commutes with time reversal operation \ie $[Q,T]=0$. Now let's add another unitary $Z_2$ symmetry $\bsc$, which is a $\pi$ spin rotation along $S^y$-axis with $\bsc^2=-1$ so that $\{Q,C\}=0$ and $[T,C]=0$. In fact $\bsq$ and $\bsc$ together generate the whole $SU(2)$ spin rotational symmetry. Now the relevant extension problem of Clifford algebra is
\bea
&\notag\{\gamma_1,\cdots,\gamma_d,TC,TQ,TQC\}\longrightarrow\\
&\{\gamma_1,\cdots,\gamma_d,TC,TQ,TQC,M\}\notag.
\eea
As a result topological singlet superconductors in $d$ dimensions are classified by $\pi_0(R_{5-d})$ for class CII ($\bst^2=+1$), and by $\pi_0(R_{-1-d})=\pi_0(R_{7-d})$ for class CI ($\bst^2=-1$).

The last example related to real Clifford algebra is class C, with a $U(1)$ symmetry $\bsq$ and a unitary $Z_2$ symmetry $\bsc$, satisfying $\bsc^2=-1$ and $\{Q,C\}=0$. As discussed earlier the system actually has a $SU(2)$ symmetry. Here we consider extension problem
\bea
\{\gamma_1Q,\cdots,\gamma_dQ,C,QC\}\rightarrow \notag\{\gamma_1Q,\cdots,\gamma_dQ,C,QC,QM\}
\eea
Notice that $(QM)^2=-M^2=+1$ and $(\gamma_iQ)^2=-1$, hence the Clifford algebra is $Cl_{0,d+2}\rightarrow Cl_{1,d+2}$. This extension problem is equivalent to the following one $Cl_{d+4,0}\rightarrow Cl_{d+4,1}$:
\bea
&\{\imth\sigma_y\otimes\gamma_iQ,\imth\sigma_y\otimes C,\imth\sigma_y\otimes QC,\sigma_x\otimes1,\sigma_z\otimes1\}\rightarrow \notag\\
&\notag\{\imth\sigma_y\otimes\gamma_iQ,\imth\sigma_y\otimes C,\imth\sigma_y\otimes QC,\sigma_x\otimes1,\sigma_z\otimes1,\imth\sigma_y\otimes QM\}
\eea
by direct product with $2\times2$ Pauli matrices $\sigma_{x,y,z}$. Therefore singlet superconductors in class C are classified by $\pi_0(R_{-2-d})=\pi_0(R_{6-d})$ in $d$ dimensions.\\

\subsection{Two complex classes}\label{app:10-fold-way:complex}

Now let's consider the case with only $U(1)$ symmetry $\bsq$. Since its generator $Q^2=-1$, without loss of generality we can always choose $Q=\imth\sigma_y\otimes1$ in a proper basis for Dirac Hamiltonian (\ref{dirac}). Since $[Q,\gamma_i]=[Q,M]=0$ we notice that Dirac matrices and the mass matrix have the following generic form
\bea
&\notag\gamma_i=1_{2\times2}\otimes\rho_{i}+\imth\sigma_y\otimes\chi_{i},~~~\rho_i^T=\rho_i,~~~\chi_i^T=-\chi_i,\\
&\notag M=1_{2\times2}\otimes M_a+\imth\sigma_y\otimes M_s,~~~M_s^T=M_s,~~~M_a^T=-M_a.
\eea
where $\rho_i,\chi_i$ and $M_a,M_s$ are all real matrices. Meanwhile conditions (\ref{cond:gamma})-(\ref{cond:mass}) indicate that Hermitian matrices
\bea
&\Gamma_i\equiv\rho_i+\imth\chi_i,~~~M_H\equiv M_s+\imth M_a.
\eea
generate a complex Clifford algebra $Cl_{d+1}$ as in (\ref{clifford:complex}). Therefore the classifying space for mass matrix $M$ (or equivalently $M_H$) is $C_d$, associated with the extension problem of complex Clifford algebra $Cl_{d}\rightarrow Cl_{d+1}$. Hence the classification of insulators in class A in $d$ spatial dimensions is given by $\pi_0(C_d)$.

In the presence of time reversal symmetry $\bst$ and $U(1)$ spin rotational symmetry $\bsq$, since $[Q,T]=0$ and $Q^2=-1$, as discussed earlier, this extra $U(1)$ symmetry rearranges the original extension problem $Cl_{d,1}\rightarrow Cl_{d,2}$ (generated by $\{\gamma_i,T\}\rightarrow\{\gamma_i,T,M\}$) into a complex Clifford algebra extension problem $Cl_{d+1}\rightarrow Cl_{d+2}$. Hence the classification of gapped superconductors in class AIII is given by $\pi_0(C_{d+1})$.

More generally, for any real Clifford algebra $Cl_{p,q}$, consider an extra symmetry ${\boldsymbol{U}}$ (such as the $U(1)$ symmetry discussed earlier) whose generator $U$ commutes with all generators of this real Clifford algebra. If $U^2=-1$, we reach a similar conclusion that the original extension problem $Cl_{p,q}\rightarrow Cl_{p,q+1}$ now becomes an extension problem $Cl_{p+q}\rightarrow Cl_{p+q+1}$, and the classifying space for the mass matrix is $C_{p+q}$. On the other hand, if $U^2=+1$, we can always choose a proper basis so that
symmetry ${\boldsymbol{U}}$ is diagonalized as $U=\sigma_z\otimes1$. Therefore we can work in two subspaces with $U=+1$ and $U=-1$ independently, since they will never mix for non-interacting fermions.

\section{Gapped surface of inversion-protected 3d singlet superconductor}\label{app:surface:class C:3d}

In this section we derive the gapped surface states of inversion-protected $\nu=1$ singlet superconductor in 3d (class C in TABLE \ref{tab:10-fold way}), based on effective Dirac Hamiltonian (\ref{dirac:C:3d}). The ``half'' spin quantum Hall effect on the surface is also obtained.

Consider periodic boundary condition in $\hat{x}$ and $\hat{y}$ direction, and an open surface parallel to $\hat{x}-o-\hat{y}$ plane located at $z=0$. The vacuum is in the $z<0$ area while $z>0$ corresponds to the bulk inversion-protected $\nu=1$ topological insulator. The surface state $|Surf\rangle$ is described by zero-energy solution of the following Dirac equation with a mass domain wall $m(z)=|m(z)|\cdot\text{Sgn}(z)$
\bea
\big[-\imth\partial_z\cdot\sigma_z\tau_y+m(z)\tau_x\mu_y\big]|Surf\rangle=0
\eea
which is
\bea
|Surf\rangle\sim e^{-\int_0^zm(\lambda)\text{d}\lambda}|\sigma_z\tau_z\mu_y=+1\rangle.
\eea
Therefore $\mu_y=\sigma_z\tau_z$ for the low-energy states on surface, which are described by
\bea\label{dirac:C:surface}
\mathcal{H}^C_{surf}=\sum_{\bf k}\tilde\Psi_{\bf k}^\dagger\Big[(k_x\sigma_x+k_y\sigma_y)\tau_y+m_s\sigma_z\Big]\tilde\Psi_{\bf k}
\eea
where $\sigma_z$ is the only surface mass term allowed by $SU(2)$ spin rotational symmetry. Dirac fermion basis $\tilde\Psi_{\bf k}$ is obtained by projecting original electron basis $\Psi_{\bf k}$ into the $\sigma_z\tau_z\mu_y=+1$ subspace. It's straightforward to check that surface model (\ref{dirac:C:surface}) has Chern number $C_s=\text{Sgn}(m_s)$. Correspondingly the spin quantum Hall conductance on the surface is given by\cite{Ludwig1994,Senthil1999,Read2000}
\bea
\sigma_{xy}^{\text{spin}}=C_s\frac{(\hbar/2)^2}{h}=C_s\cdot\frac{\hbar}{8\pi},~~~C_s=\pm1.
\eea
Note that in any two-dimensional singlet superconductors (such as $d+\imth d$ superconductor) the spin Hall conductance $\sigma_{xy}^{\text{spin}}$ is always an \emph{even}
integer\cite{Senthil1999} in unit of $\hbar/8\pi$. Therefore we say the gapped surface states of inversion-protected $\nu=1$ singlet (class C) superconductor exhibits ``half'' spin quantum Hall effect.


\begin{thebibliography}{37}
\expandafter\ifx\csname natexlab\endcsname\relax\def\natexlab#1{#1}\fi
\expandafter\ifx\csname bibnamefont\endcsname\relax
  \def\bibnamefont#1{#1}\fi
\expandafter\ifx\csname bibfnamefont\endcsname\relax
  \def\bibfnamefont#1{#1}\fi
\expandafter\ifx\csname citenamefont\endcsname\relax
  \def\citenamefont#1{#1}\fi
\expandafter\ifx\csname url\endcsname\relax
  \def\url#1{\texttt{#1}}\fi
\expandafter\ifx\csname urlprefix\endcsname\relax\def\urlprefix{URL }\fi
\providecommand{\bibinfo}[2]{#2}
\providecommand{\eprint}[2][]{\url{#2}}

\bibitem[{\citenamefont{Hasan and Kane}(2010)}]{Hasan2010}
\bibinfo{author}{\bibfnamefont{M.~Z.} \bibnamefont{Hasan}} \bibnamefont{and}
  \bibinfo{author}{\bibfnamefont{C.~L.} \bibnamefont{Kane}},
  \bibinfo{journal}{Rev. Mod. Phys.} \textbf{\bibinfo{volume}{82}},
  \bibinfo{pages}{3045} (\bibinfo{year}{2010}).

\bibitem[{\citenamefont{Hasan and Moore}(2011)}]{Hasan2011}
\bibinfo{author}{\bibfnamefont{M.~Z.} \bibnamefont{Hasan}} \bibnamefont{and}
  \bibinfo{author}{\bibfnamefont{J.~E.} \bibnamefont{Moore}},
  \bibinfo{journal}{Annu. Rev. Condens. Matter Phys.}
  \textbf{\bibinfo{volume}{2}}, \bibinfo{pages}{55} (\bibinfo{year}{2011}),
  ISSN \bibinfo{issn}{1947-5454}.

\bibitem[{\citenamefont{Qi and Zhang}(2011)}]{Qi2011}
\bibinfo{author}{\bibfnamefont{X.-L.} \bibnamefont{Qi}} \bibnamefont{and}
  \bibinfo{author}{\bibfnamefont{S.-C.} \bibnamefont{Zhang}},
  \bibinfo{journal}{Rev. Mod. Phys.} \textbf{\bibinfo{volume}{83}},
  \bibinfo{pages}{1057} (\bibinfo{year}{2011}).

\bibitem[{\citenamefont{Chen et~al.}(2013)\citenamefont{Chen, Gu, Liu, and
  Wen}}]{Chen2013}
\bibinfo{author}{\bibfnamefont{X.}~\bibnamefont{Chen}},
  \bibinfo{author}{\bibfnamefont{Z.-C.} \bibnamefont{Gu}},
  \bibinfo{author}{\bibfnamefont{Z.-X.} \bibnamefont{Liu}}, \bibnamefont{and}
  \bibinfo{author}{\bibfnamefont{X.-G.} \bibnamefont{Wen}},
  \bibinfo{journal}{Phys. Rev. B} \textbf{\bibinfo{volume}{87}},
  \bibinfo{pages}{155114} (\bibinfo{year}{2013}).

\bibitem[{\citenamefont{Fu et~al.}(2007)\citenamefont{Fu, Kane, and
  Mele}}]{Fu2007}
\bibinfo{author}{\bibfnamefont{L.}~\bibnamefont{Fu}},
  \bibinfo{author}{\bibfnamefont{C.~L.} \bibnamefont{Kane}}, \bibnamefont{and}
  \bibinfo{author}{\bibfnamefont{E.~J.} \bibnamefont{Mele}},
  \bibinfo{journal}{Phys. Rev. Lett.} \textbf{\bibinfo{volume}{98}},
  \bibinfo{pages}{106803} (\bibinfo{year}{2007}).

\bibitem[{\citenamefont{Moore and Balents}(2007)}]{Moore2007}
\bibinfo{author}{\bibfnamefont{J.~E.} \bibnamefont{Moore}} \bibnamefont{and}
  \bibinfo{author}{\bibfnamefont{L.}~\bibnamefont{Balents}},
  \bibinfo{journal}{Phys. Rev. B} \textbf{\bibinfo{volume}{75}},
  \bibinfo{pages}{121306} (\bibinfo{year}{2007}).

\bibitem[{\citenamefont{Roy}(2009)}]{Roy2009}
\bibinfo{author}{\bibfnamefont{R.}~\bibnamefont{Roy}}, \bibinfo{journal}{Phys.
  Rev. B} \textbf{\bibinfo{volume}{79}}, \bibinfo{pages}{195322}
  (\bibinfo{year}{2009}).

\bibitem[{\citenamefont{Essin et~al.}(2009)\citenamefont{Essin, Moore, and
  Vanderbilt}}]{Essin2009}
\bibinfo{author}{\bibfnamefont{A.~M.} \bibnamefont{Essin}},
  \bibinfo{author}{\bibfnamefont{J.~E.} \bibnamefont{Moore}}, \bibnamefont{and}
  \bibinfo{author}{\bibfnamefont{D.}~\bibnamefont{Vanderbilt}},
  \bibinfo{journal}{Phys. Rev. Lett.} \textbf{\bibinfo{volume}{102}},
  \bibinfo{pages}{146805} (\bibinfo{year}{2009}).

\bibitem[{\citenamefont{Kitaev}(2009)}]{Kitaev2009}
\bibinfo{author}{\bibfnamefont{A.}~\bibnamefont{Kitaev}}, \bibinfo{journal}{AIP
  Conf. Proc.} \textbf{\bibinfo{volume}{1134}}, \bibinfo{pages}{22}
  (\bibinfo{year}{2009}).

\bibitem[{\citenamefont{Teo and Kane}(2010)}]{Teo2010}
\bibinfo{author}{\bibfnamefont{J.~C.~Y.} \bibnamefont{Teo}} \bibnamefont{and}
  \bibinfo{author}{\bibfnamefont{C.~L.} \bibnamefont{Kane}},
  \bibinfo{journal}{Phys. Rev. B} \textbf{\bibinfo{volume}{82}},
  \bibinfo{pages}{115120} (\bibinfo{year}{2010}).

\bibitem[{\citenamefont{Schnyder et~al.}(2008)\citenamefont{Schnyder, Ryu,
  Furusaki, and Ludwig}}]{Schnyder2008}
\bibinfo{author}{\bibfnamefont{A.~P.} \bibnamefont{Schnyder}},
  \bibinfo{author}{\bibfnamefont{S.}~\bibnamefont{Ryu}},
  \bibinfo{author}{\bibfnamefont{A.}~\bibnamefont{Furusaki}}, \bibnamefont{and}
  \bibinfo{author}{\bibfnamefont{A.~W.~W.} \bibnamefont{Ludwig}},
  \bibinfo{journal}{Phys. Rev. B} \textbf{\bibinfo{volume}{78}},
  \bibinfo{pages}{195125} (\bibinfo{year}{2008}).

\bibitem[{\citenamefont{Ran}(2010)}]{Ran2010}
\bibinfo{author}{\bibfnamefont{Y.}~\bibnamefont{Ran}}, \bibinfo{journal}{ArXiv
  e-prints 1006.5454}  (\bibinfo{year}{2010}), \eprint{1006.5454}.

\bibitem[{\citenamefont{Zhang et~al.}(2013)\citenamefont{Zhang, Kane, and
  Mele}}]{Zhang2013a}
\bibinfo{author}{\bibfnamefont{F.}~\bibnamefont{Zhang}},
  \bibinfo{author}{\bibfnamefont{C.~L.} \bibnamefont{Kane}}, \bibnamefont{and}
  \bibinfo{author}{\bibfnamefont{E.~J.} \bibnamefont{Mele}},
  \bibinfo{journal}{Phys. Rev. Lett.} \textbf{\bibinfo{volume}{111}},
  \bibinfo{pages}{056403} (\bibinfo{year}{2013}).

\bibitem[{\citenamefont{Chiu et~al.}(2013)\citenamefont{Chiu, Yao, and
  Ryu}}]{Chiu2013}
\bibinfo{author}{\bibfnamefont{C.-K.} \bibnamefont{Chiu}},
  \bibinfo{author}{\bibfnamefont{H.}~\bibnamefont{Yao}}, \bibnamefont{and}
  \bibinfo{author}{\bibfnamefont{S.}~\bibnamefont{Ryu}},
  \bibinfo{journal}{Phys. Rev. B} \textbf{\bibinfo{volume}{88}},
  \bibinfo{pages}{075142} (\bibinfo{year}{2013}).

\bibitem[{\citenamefont{Morimoto and Furusaki}(2013)}]{Morimoto2013}
\bibinfo{author}{\bibfnamefont{T.}~\bibnamefont{Morimoto}} \bibnamefont{and}
  \bibinfo{author}{\bibfnamefont{A.}~\bibnamefont{Furusaki}},
  \bibinfo{journal}{Phys. Rev. B} \textbf{\bibinfo{volume}{88}},
  \bibinfo{pages}{125129} (\bibinfo{year}{2013}).

\bibitem[{\citenamefont{{Lu} et~al.}(2014)\citenamefont{{Lu}, {Cho}, and
  {Vishwanath}}}]{Lu2014}
\bibinfo{author}{\bibfnamefont{Y.-M.} \bibnamefont{{Lu}}},
  \bibinfo{author}{\bibfnamefont{G.~Y.} \bibnamefont{{Cho}}}, \bibnamefont{and}
  \bibinfo{author}{\bibfnamefont{A.}~\bibnamefont{{Vishwanath}}},
  \bibinfo{journal}{ArXiv e-prints}  (\bibinfo{year}{2014}),
  \eprint{1403.0575}.

\bibitem[{\citenamefont{Fang et~al.}(2012)\citenamefont{Fang, Gilbert, and
  Bernevig}}]{Fang2012}
\bibinfo{author}{\bibfnamefont{C.}~\bibnamefont{Fang}},
  \bibinfo{author}{\bibfnamefont{M.~J.} \bibnamefont{Gilbert}},
  \bibnamefont{and} \bibinfo{author}{\bibfnamefont{B.~A.}
  \bibnamefont{Bernevig}}, \bibinfo{journal}{Phys. Rev. B}
  \textbf{\bibinfo{volume}{86}}, \bibinfo{pages}{115112}
  (\bibinfo{year}{2012}).

\bibitem[{\citenamefont{Teo and Hughes}(2013)}]{Teo2013}
\bibinfo{author}{\bibfnamefont{J.~C.~Y.} \bibnamefont{Teo}} \bibnamefont{and}
  \bibinfo{author}{\bibfnamefont{T.~L.} \bibnamefont{Hughes}},
  \bibinfo{journal}{Phys. Rev. Lett.} \textbf{\bibinfo{volume}{111}},
  \bibinfo{pages}{047006} (\bibinfo{year}{2013}).

\bibitem[{\citenamefont{{Benalcazar} et~al.}(2013)\citenamefont{{Benalcazar},
  {Teo}, and {Hughes}}}]{Benalcazar2013}
\bibinfo{author}{\bibfnamefont{W.~A.} \bibnamefont{{Benalcazar}}},
  \bibinfo{author}{\bibfnamefont{J.~C.~Y.} \bibnamefont{{Teo}}},
  \bibnamefont{and} \bibinfo{author}{\bibfnamefont{T.~L.}
  \bibnamefont{{Hughes}}}, \bibinfo{journal}{ArXiv e-prints}
  (\bibinfo{year}{2013}), \eprint{1311.0496}.

\bibitem[{\citenamefont{Turner et~al.}(2010)\citenamefont{Turner, Zhang, and
  Vishwanath}}]{Turner2010}
\bibinfo{author}{\bibfnamefont{A.~M.} \bibnamefont{Turner}},
  \bibinfo{author}{\bibfnamefont{Y.}~\bibnamefont{Zhang}}, \bibnamefont{and}
  \bibinfo{author}{\bibfnamefont{A.}~\bibnamefont{Vishwanath}},
  \bibinfo{journal}{Phys. Rev. B} \textbf{\bibinfo{volume}{82}},
  \bibinfo{pages}{241102} (\bibinfo{year}{2010}).

\bibitem[{\citenamefont{Wen}(2002)}]{Wen2002}
\bibinfo{author}{\bibfnamefont{X.-G.} \bibnamefont{Wen}},
  \bibinfo{journal}{Phys. Rev. B} \textbf{\bibinfo{volume}{65}},
  \bibinfo{pages}{165113} (\bibinfo{year}{2002}).

\bibitem[{\citenamefont{Wen}(2012)}]{Wen2012}
\bibinfo{author}{\bibfnamefont{X.-G.} \bibnamefont{Wen}},
  \bibinfo{journal}{Phys. Rev. B} \textbf{\bibinfo{volume}{85}},
  \bibinfo{pages}{085103} (\bibinfo{year}{2012}).

\bibitem[{\citenamefont{Thouless et~al.}(1982)\citenamefont{Thouless, Kohmoto,
  Nightingale, and den Nijs}}]{Thouless1982}
\bibinfo{author}{\bibfnamefont{D.~J.} \bibnamefont{Thouless}},
  \bibinfo{author}{\bibfnamefont{M.}~\bibnamefont{Kohmoto}},
  \bibinfo{author}{\bibfnamefont{M.~P.} \bibnamefont{Nightingale}},
  \bibnamefont{and} \bibinfo{author}{\bibfnamefont{M.}~\bibnamefont{den Nijs}},
  \bibinfo{journal}{Phys. Rev. Lett.} \textbf{\bibinfo{volume}{49}},
  \bibinfo{pages}{405} (\bibinfo{year}{1982}).

\bibitem[{\citenamefont{Kane and Mele}(2005{\natexlab{a}})}]{Kane2005}
\bibinfo{author}{\bibfnamefont{C.~L.} \bibnamefont{Kane}} \bibnamefont{and}
  \bibinfo{author}{\bibfnamefont{E.~J.} \bibnamefont{Mele}},
  \bibinfo{journal}{Phys. Rev. Lett.} \textbf{\bibinfo{volume}{95}},
  \bibinfo{pages}{226801} (\bibinfo{year}{2005}{\natexlab{a}}).

\bibitem[{\citenamefont{Kane and Mele}(2005{\natexlab{b}})}]{Kane2005a}
\bibinfo{author}{\bibfnamefont{C.~L.} \bibnamefont{Kane}} \bibnamefont{and}
  \bibinfo{author}{\bibfnamefont{E.~J.} \bibnamefont{Mele}},
  \bibinfo{journal}{Phys. Rev. Lett.} \textbf{\bibinfo{volume}{95}},
  \bibinfo{pages}{146802} (\bibinfo{year}{2005}{\natexlab{b}}).

\bibitem[{\citenamefont{Bernevig and Zhang}(2006)}]{Bernevig2006}
\bibinfo{author}{\bibfnamefont{B.~A.} \bibnamefont{Bernevig}} \bibnamefont{and}
  \bibinfo{author}{\bibfnamefont{S.-C.} \bibnamefont{Zhang}},
  \bibinfo{journal}{Phys. Rev. Lett.} \textbf{\bibinfo{volume}{96}},
  \bibinfo{pages}{106802} (\bibinfo{year}{2006}).

\bibitem[{\citenamefont{Read and Green}(2000)}]{Read2000}
\bibinfo{author}{\bibfnamefont{N.}~\bibnamefont{Read}} \bibnamefont{and}
  \bibinfo{author}{\bibfnamefont{D.}~\bibnamefont{Green}},
  \bibinfo{journal}{Phys. Rev. B} \textbf{\bibinfo{volume}{61}},
  \bibinfo{pages}{10267} (\bibinfo{year}{2000}).

\bibitem[{\citenamefont{Hughes et~al.}(2011)\citenamefont{Hughes, Prodan, and
  Bernevig}}]{Hughes2011}
\bibinfo{author}{\bibfnamefont{T.~L.} \bibnamefont{Hughes}},
  \bibinfo{author}{\bibfnamefont{E.}~\bibnamefont{Prodan}}, \bibnamefont{and}
  \bibinfo{author}{\bibfnamefont{B.~A.} \bibnamefont{Bernevig}},
  \bibinfo{journal}{Phys. Rev. B} \textbf{\bibinfo{volume}{83}},
  \bibinfo{pages}{245132} (\bibinfo{year}{2011}).

\bibitem[{\citenamefont{Turner et~al.}(2012)\citenamefont{Turner, Zhang, Mong,
  and Vishwanath}}]{Turner2012}
\bibinfo{author}{\bibfnamefont{A.~M.} \bibnamefont{Turner}},
  \bibinfo{author}{\bibfnamefont{Y.}~\bibnamefont{Zhang}},
  \bibinfo{author}{\bibfnamefont{R.~S.~K.} \bibnamefont{Mong}},
  \bibnamefont{and}
  \bibinfo{author}{\bibfnamefont{A.}~\bibnamefont{Vishwanath}},
  \bibinfo{journal}{Phys. Rev. B} \textbf{\bibinfo{volume}{85}},
  \bibinfo{pages}{165120} (\bibinfo{year}{2012}).

\bibitem[{\citenamefont{Qi et~al.}(2008)\citenamefont{Qi, Hughes, and
  Zhang}}]{Qi2008}
\bibinfo{author}{\bibfnamefont{X.-L.} \bibnamefont{Qi}},
  \bibinfo{author}{\bibfnamefont{T.~L.} \bibnamefont{Hughes}},
  \bibnamefont{and} \bibinfo{author}{\bibfnamefont{S.-C.} \bibnamefont{Zhang}},
  \bibinfo{journal}{Phys. Rev. B} \textbf{\bibinfo{volume}{78}},
  \bibinfo{pages}{195424} (\bibinfo{year}{2008}).

\bibitem[{\citenamefont{Senthil et~al.}(1999)\citenamefont{Senthil, Marston,
  and Fisher}}]{Senthil1999}
\bibinfo{author}{\bibfnamefont{T.}~\bibnamefont{Senthil}},
  \bibinfo{author}{\bibfnamefont{J.~B.} \bibnamefont{Marston}},
  \bibnamefont{and} \bibinfo{author}{\bibfnamefont{M.~P.~A.}
  \bibnamefont{Fisher}}, \bibinfo{journal}{Phys. Rev. B}
  \textbf{\bibinfo{volume}{60}}, \bibinfo{pages}{4245} (\bibinfo{year}{1999}).

\bibitem[{\citenamefont{Schnyder et~al.}(2009)\citenamefont{Schnyder, Ryu, and
  Ludwig}}]{Schnyder2009}
\bibinfo{author}{\bibfnamefont{A.~P.} \bibnamefont{Schnyder}},
  \bibinfo{author}{\bibfnamefont{S.}~\bibnamefont{Ryu}}, \bibnamefont{and}
  \bibinfo{author}{\bibfnamefont{A.~W.~W.} \bibnamefont{Ludwig}},
  \bibinfo{journal}{Phys. Rev. Lett.} \textbf{\bibinfo{volume}{102}},
  \bibinfo{pages}{196804} (\bibinfo{year}{2009}).

\bibitem[{\citenamefont{Fang et~al.}(2013)\citenamefont{Fang, Gilbert, and
  Bernevig}}]{Fang2013a}
\bibinfo{author}{\bibfnamefont{C.}~\bibnamefont{Fang}},
  \bibinfo{author}{\bibfnamefont{M.~J.} \bibnamefont{Gilbert}},
  \bibnamefont{and} \bibinfo{author}{\bibfnamefont{B.~A.}
  \bibnamefont{Bernevig}}, \bibinfo{journal}{Phys. Rev. B}
  \textbf{\bibinfo{volume}{87}}, \bibinfo{pages}{035119}
  (\bibinfo{year}{2013}).

\bibitem[{\citenamefont{Fidkowski and Kitaev}(2010)}]{Fidkowski2010}
\bibinfo{author}{\bibfnamefont{L.}~\bibnamefont{Fidkowski}} \bibnamefont{and}
  \bibinfo{author}{\bibfnamefont{A.}~\bibnamefont{Kitaev}},
  \bibinfo{journal}{Phys. Rev. B} \textbf{\bibinfo{volume}{81}},
  \bibinfo{pages}{134509} (\bibinfo{year}{2010}).

\bibitem[{\citenamefont{Turner et~al.}(2011)\citenamefont{Turner, Pollmann, and
  Berg}}]{Turner2011}
\bibinfo{author}{\bibfnamefont{A.~M.} \bibnamefont{Turner}},
  \bibinfo{author}{\bibfnamefont{F.}~\bibnamefont{Pollmann}}, \bibnamefont{and}
  \bibinfo{author}{\bibfnamefont{E.}~\bibnamefont{Berg}},
  \bibinfo{journal}{Phys. Rev. B} \textbf{\bibinfo{volume}{83}},
  \bibinfo{pages}{075102} (\bibinfo{year}{2011}).

\bibitem[{\citenamefont{{Shiozaki} and {Sato}}(2014)}]{Shiozaki2014}
\bibinfo{author}{\bibfnamefont{K.}~\bibnamefont{{Shiozaki}}} \bibnamefont{and}
  \bibinfo{author}{\bibfnamefont{M.}~\bibnamefont{{Sato}}},
  \bibinfo{journal}{ArXiv e-prints}  (\bibinfo{year}{2014}),
  \eprint{1403.3331}.

\bibitem[{\citenamefont{Ludwig et~al.}(1994)\citenamefont{Ludwig, Fisher,
  Shankar, and Grinstein}}]{Ludwig1994}
\bibinfo{author}{\bibfnamefont{A.~W.~W.} \bibnamefont{Ludwig}},
  \bibinfo{author}{\bibfnamefont{M.~P.~A.} \bibnamefont{Fisher}},
  \bibinfo{author}{\bibfnamefont{R.}~\bibnamefont{Shankar}}, \bibnamefont{and}
  \bibinfo{author}{\bibfnamefont{G.}~\bibnamefont{Grinstein}},
  \bibinfo{journal}{Phys. Rev. B} \textbf{\bibinfo{volume}{50}},
  \bibinfo{pages}{7526} (\bibinfo{year}{1994}).

\end{thebibliography}

\end{document}